\documentclass[twocolumn]{autart}    

\usepackage{amssymb,amsmath,amsfonts}
\usepackage{graphicx}
\usepackage{epstopdf}
\usepackage{color}
\usepackage{enumitem}
\usepackage{soul}
\usepackage[authoryear,round]{natbib}

\newcommand*{\red}{\textcolor{black}}

\newcommand{\Sc}{\mathcal{S}}

\newcommand{\Zc}{\mathcal{Z}}
\newcommand{\Lc}{\mathcal{L}}
\newcommand{\Jc}{\mathcal{J}}

\newcommand{\Vc}{\mathcal{V}}

\newcommand{\Kc}{\mathcal{K}}

\newcommand{\tSc}{\tilde{\mathcal{S}}}

\newcommand{\tZc}{\tilde{\mathcal{Z}}}

\newcommand{\R}{\mathbb{R}}

\begin{document}

\begin{frontmatter}

\title{Abstractions of linear dynamic networks for input selection in local module
identification\thanksref{footnoteinfo}} 

\thanks[footnoteinfo]{Paper accepted for publication in Automatica. Submitted 2 January 2019, revised 12 September 2019, final version 16 March 2020. This project has received funding from the European Research Council (ERC), Advanced Research Grant SYSDYNET, under the European Union's Horizon 2020 research and innovation programme (Grant Agreement No. 694504), and the Vinnova Industry Excellence Center LINK-SIC project number 2007-02224.}

\author[First]{Harm H.M. Weerts},
\author[Second]{Jonas Linder},
\author[Third]{Martin Enqvist} and
\author[First]{Paul M.J. Van den Hof}
\address[First]{Control Systems Group, Department of Electrical Engineering, Eindhoven University of Technology, The Netherlands (email: harmweerts@gmail.com, p.m.j.vandenhof@tue.nl)}
\address[Second]{ABB Corporate Research, V\"{a}ster{\aa}s, Sweden (e-mail: jonas.x.linder@se.abb.com)}
\address[Third]{Division of Automatic Control, Link\"{o}ping University, Sweden (e-mail: martin.enqvist@liu.se)}


\begin{abstract}                          
In abstractions of linear dynamic networks, selected node signals are removed from the network, while keeping the remaining node signals invariant.
The topology and link dynamics, or modules, of an abstracted network will generally be changed compared to the original network.
Abstractions of dynamic networks can be used to select an appropriate set of node signals that are to be measured, on the basis of which a particular local module can be estimated.
A method is introduced for network abstraction that generalizes previously introduced algorithms, as e.g. immersion and the method of indirect inputs.
For this abstraction method it is shown under which conditions on the selected signals a particular module will remain invariant.
This leads to sets of conditions on selected measured node variables that allow identification of the target module.
%
\end{abstract}

\end{frontmatter}

\section{Introduction}
%
In current and future engineering systems, as well as in many biological and biomedical systems, large scale interconnected dynamical systems become a prime modelling target for analysis and control. The resulting dynamic networks that represent these interconnected systems, are studied from many different perspectives. In this paper our main motivation is directed towards methods and tools for data-driven modelling of (parts of) a dynamic network, extending identification methods to be able to deal with linear dynamic networks.

In network identification literature, typically three main objectives can be distinguished.
One objective is to perform identification of both the topology and dynamics, see for example \cite{Sanandaji2011,ChiusoPAuto2012,MaterassiSalapaka2012,Zorzi&Chiuso:17}.
In situations where there is prior knowledge of the topology, the objective can be to identify all dynamic modules, as is done for example in  \cite{Goncalves08,Yuan2011,weerts2018rr}, including addressing the aspect of network identifiability, \citep{weerts2018identifiability,Hendrickx&Gevers&Bazanella_TAC:19}.
A third possible objective is the identification of the dynamics of a single local element, or module, in the network, when the topology of the network is known. This has been addressed in \cite{VandenHof&Dankers&etal:13}, by extending classical  closed-loop prediction error methods, known as the direct method, two-stage method and joint-io method, to the situation of dynamic networks. The reasoning has been extended in \cite{dankers2015errors} to deal with signal measurements that are affected by sensor noise. Additional extensions concern identifiability aspects of local modules \citep{bazanella2017identifiability,Hendrickx&Gevers&Bazanella_TAC:19,Gevers&etal_sysid:18,weerts2018_cdc}, and combination of the prediction error method with Bayesian estimation in \cite{Everitt:17b,ramaswamy2018}.

In this paper we will elaborate on this local module identification problem, and focus on the selection of signals to be measured in the network,  to allow a particular module to be identified. While in general it may be attractive to have all node signals in a network available from measurements, in practice it may be costly or even impossible to measure some of the nodes.
Then, a relevant question is which set of measured node signals is
sufficient for single module identification. This problem has been addressed in several different ways.
\begin{itemize}[leftmargin=*]
	\item
	In \cite{dankers2016identification} the module of interest is identified in a multi-input-single-output setting with either a direct method or a two-stage method.
Signals are selected by removing non-selected signals from the network through an elimination procedure called {\it immersion}, while requiring that the target module remains invariant. This allows a consistent identification of the target module dynamics under appropriate excitation and disturbance conditions. A systematic way of selecting the measured node signals is provided. The selection method is extended in \cite{dankers2017conditions} to handle situations of confounding variables being caused by disturbances that are spatially correlated over the different node signals. This is particularly of interest when applying the direct method, aiming at maximum likelihood results.
	\item
	The approach in  \cite{linder2017,linder2017ifac,linder2017identification}  is also in a  multi-input-single-output setting, but with an instrumental variable identification method. It follows the same philosophy as the one in \cite{dankers2016identification}, but applies a different elimination procedure, referred to as the \emph{indirect inputs method}, and thus results in a different set of conditions on selected measured node signals.
\item In \cite{materassi2015identification,Materassi&Salapaka:19}, selection of node signals is being done on the basis of graphical models, while Wiener filters are being used for dynamic model reconstruction, and no reference excitations are present.
This approach can be characterized as an indirect approach where several dynamic objects are identified from data, after which the target module is reconstructed.
\item
	The approach in \cite{bazanella2017identifiability,Hendrickx&Gevers&Bazanella_TAC:19} is also based on indirect identification for the situation that all nodes are excited by external excitations.
	Here an identifiability analysis is made in order to verify whether the module of interest can be recovered uniquely from a set of estimated transfer functions from reference signals to a set of selected node signals.
\end{itemize}
Since all algorithms provide only sufficient conditions for arriving at a set of to-be-measured node signals, it is attractive to obtain results that fully characterize the degrees of freedom that are available in selecting node signals that allow appropriate module estimates.

In this paper we are going to adopt the strategy of the first two approaches, eliminating node variables to arrive at a so-called {\it abstracted network}, while keeping the dynamics of the target module invariant.
After abstraction, the target module can be estimated on the basis of the node variables that are retained in the abstraction.
While both \cite{dankers2016identification} and \cite{linder2017ifac} have employed a particular way of abstracting networks, e.g. through network immersion (Kron reduction) or through the so-called indirect inputs method, here we develop a more general notion of abstraction that generalizes the two earlier approaches, and provides a higher flexibility in the selection mechanism of choosing which nodes to measure/retain and which nodes to discard/abstract in a multi-input single-output identification setup for estimating the target module.

In order to develop this generalized abstraction algorithm, we use the fundamental property that
the network representation is not a unique representation of the behavior of the network.
When manipulating the network equations, different representations can be obtained leading to different identification setups for estimating the target module.
This freedom will be exploited for developing generalized conditions for selecting the node signals that are used as inputs in an identification setup for estimating the target module.

The paper is organized as follows. After defining the basis network setup in Section \ref{sec:netw_def}, in Section \ref{sect:equiv} the non-uniqueness of the network representations is characterized.
This is exploited in Section \ref{sect:abstract} to arrive at a generalized abstraction algorithm, which is illustrated by examples in Section \ref{sec:exam}.
In Section \ref{sect:invariance} it is specified under which conditions the target module remains invariant, and consequences for the identification setup and the selection of node signals are presented in Section \ref{sect:identsetup}. The proofs of all results are collected in the Appendix.
\section{Dynamic network definition}  \label{sec:netw_def}
In this section a dynamic network model is formulated on the basis of the setup in \cite{VandenHof&Dankers&etal:13}.
A dynamic network consists of $L$ scalar \emph{internal variables} or \emph{nodes} $w_j$, $j =1,\ldots,L$, and $K$ \emph{external variables} $r_k$, $k=1,\ldots,K$.
Each node is a basic building block of the network and is described as:
\begin{align}
w_j(t) = \sum_{\substack{l=1 \\ l\neq j}}^L
G_{jl}(q)w_l(t) + u_j(t) + v_j(t)
\label{eq:netw_def}
\end{align}
where $q^{-1}$ is the delay operator, i.e. $q^{-1}w_j(t) = w_j(t-1)$;
\begin{itemize}[leftmargin=*]
	\item $G_{jl}$ are proper rational transfer functions that are referred to as {\it modules} in the network;
	\item There are no self-loops in the network, i.e. nodes are not directly connected to themselves $G_{jj} = 0$;\footnote{Since $G_{jj}$ are rational transfer functions, this does not limit the dynamical description of $w_j(t)$.}
	\item  $u_j(t)$ are generated by the \emph{external variables} $r_k(t)$ via
	\begin{equation} u_j(t) = \sum_{k=1}^K R_{jk}(q)r_k(t), \label{eq:Rdef} \end{equation} where $r_k$  can directly be manipulated by the user,
	and $R_{jk}$ are proper rational transfer functions;
	\item  $v_j$ is \emph{process noise}, where the vector process $v=[v_1 \cdots v_L]^T$ is modeled as a stationary stochastic process with rational spectral density, such that there exists an $L$-dimensional white noise process $e:= [e_1 \cdots e_L]^T$, with covariance matrix $\Lambda>0$ such that
\[ v(t) = H(q)e(t), \]
with $H(q)$ a proper rational transfer function.
\end{itemize}
By combining the $L$ node signals, the network expression
\begin{align}
	\begin{bmatrix}  \!  w_1  \!    \\[1pt]  \!  w_2   \!   \\[1pt]   \vdots   \\[1pt]   \!  w_L    \!  \end{bmatrix}      \! \!  =    \!   \!
	\begin{bmatrix}
		0 &  G_{12}  &   \cdots   &   G_{1L}  \\
  		 \! G_{21}  & 0 &   \ddots   &    \vdots  \\
		\vdots &  \ddots  &   \ddots   &   G_{L-1 \ L}  \!  \\
  		 \! G_{L1}  &  \cdots  &    G_{L \ L-1}    &   0
		\end{bmatrix}    \!   \!  \!
	\begin{bmatrix}   w_1  \\[1pt]    w_2  \\[1pt]   \vdots  \\[1pt]   w_L   \end{bmatrix}
	 \!  \! +  \!   \!
	\begin{bmatrix}   u_1  \\[1pt]   u_2  \\[1pt]   \vdots  \\[1pt]    u_{L}  \end{bmatrix}
	 \! \!  + \!   \!
	\begin{bmatrix}  v_1  \\[1pt]   v_2  \\[1pt]   \vdots  \\[1pt]   v_L \end{bmatrix} \!
\end{align}
is obtained,
where the zeros are due to absence of self-loops.
In  matrix notation the dynamic network is represented as
\begin{align} \label{eq.dgsMatrix}
	w(t) = G(q) w(t) + R(q) r(t) + H(q) e(t).
\end{align}
A set notation is introduced for notational convenience.
Let the sets $\Sc$ and $\Zc$ each contain a number of node indices, then $w_\Sc$ denotes the vector of node signals consisting of all $w_j, j \in \Sc$.
Similarly, $G_{\Sc \Zc}$ is the matrix of transfer functions that contains all modules $G_{ji}, j \in \Sc, i \in \Zc$.

The  transfer function that maps the external signals $r$ and $e$ into the node signals $w$ is denoted by:
\begin{equation}
	\label{eq:T0}
	T(q) = \begin{bmatrix} T_{wr}(q) & T_{we}(q) \end{bmatrix},
\end{equation}
with
\begin{align}
	&T_{wr}(q) := \left( I-G(q)\right)^{-1}R(q), \ \mbox{and}& \label{eq:twr}\\
	&T_{we}(q) := \left( I-G(q)\right)^{-1}H(q).& \label{eq:twe}
\end{align}
This is also known as the open-loop response of the network corresponding with
\begin{equation} \label{eq:ol_barv}
 	w(t) = T_{wr} r(t) + \bar{v}(t),
\end{equation}
where noise component $\bar v(t)$ is defined by
\begin{equation}
	\bar v(t) := T_{we}(q) e(t),
\end{equation}
with power spectral density
\begin{equation} \label{eq:vbar}
	\Phi_{\bar v}(\omega) := T_{we}(e^{i\omega}) \Lambda T_{we}^{T}(e^{-i\omega}).
\end{equation}
Some notions from graph theory will be used in the dynamic network.
Modules form the interconnections / links between nodes.
A node $w_k$ is said to be an \emph{in-neighbor} of node $w_j$ if  $G_{jk} \ne 0$, and then $w_j$ is said to be an \emph{out-neighbor}  of node $w_k$.
A \emph{path} in a network is a sequence of interconnected nodes, more precisely there exists a path through nodes $w_{n_1}, \ldots, w_{n_k}$ if
\[ G_{n_1 n_2} G_{n_2 n_3} \cdots G_{n_{(k-1)} n_k} \neq 0. \]
A \emph{loop} is a path where $n_1 = n_k$.
%
%
%
%
%

A dynamic network model is then formally defined in the following way.
\begin{defn}[dynamic network model]
\label{def:netw_model}
A network model of a network with $L$ nodes, and $K$ external excitation signals, with a noise process of rank $L$ is defined by the quadruple:
\[ M = (G,R,H,\Lambda) \]
with
\begin{itemize}[leftmargin=*]
\item $G \in \R^{L \times L}(z)$, diagonal entries 0;
\item $R \in \R^{L \times K}(z)$;
\item $H \in \R^{L \times L}(z)$, monic, proper, stable, with a stable inverse;
\item $\Lambda \in \R^{L\times L}$, $\Lambda > 0$;
\item The network is well-posed \citep{Dankers_diss}, implying that $(I-G)^{-1}$ exists. Additionally we require that $(I-G)^{-1}\red{R}$ is proper and stable. \hfill $\Box$
\end{itemize}
\end{defn}
\begin{rem}
	In Definition \ref{def:netw_model} we do not require modules to be proper, while in \eqref{eq:netw_def} we do require properness.
	Real systems in practice are reflected by \eqref{eq:netw_def}, and these will typically have proper modules.
	In this paper we will use the freedom to allow network models to have non-proper modules.
	For this reason we define a representation of these networks that allows modules to be non-proper in  Definition \ref{def:netw_model}. Note that for a network model representation as in Definition \ref{def:netw_model}, we require $T_{wr}$ to be proper and stable, while the noise transfer $T_{we}$ is allowed to be non-proper, representing a non-causal mapping. This allows us to maintain a monic, proper, stable and stably invertible noise filter $H$, which is attractive from an identification perspective.
\end{rem}
\begin{rem}
    With the analysis provided in \cite{weerts2018rr}, it is possible to extend the results in the current paper to the situation of a noise process having rank $p$ smaller than $L$, implying that $H \in \R^{L\times p}(z)$, and $\Lambda \in \R^{p\times p}$.
\end{rem}
\section{Equivalent network representations} \label{sect:equiv}
The freedom that is present in dynamic network representations allows for different selections of node signals to be used for identification of a module.
This freedom is formally characterized in this section.
Moreover, the general concept of removing a node from a network is defined as abstraction, such that in later sections we can consider abstractions that are relevant for identification.
\subsection{Transformation of the global network}
Fundamentally, we need to define when two networks are equivalent descriptions of behavior, and what freedom is available to transform the network to an equivalent representation.
In the network model definition, it has been stated that the external variables $r$ and nodes $w$ are known, and it is reasonable to state that equivalent networks must describe the same relation between $r$ and $w$.
The dynamic influence of $r$ on $w$ is described by the open-loop transfer function matrix $T_{wr}$, and so the equivalence of two networks additionally requires equality of the two related open-loop transfer function matrices from $r$ to $w$.
The open-loop response of the network is described by (\ref{eq:ol_barv}), i.e. $w(t) = T_{wr} r(t) + \bar{v}(t)$.
If $w$, $r$ and $T_{wr}$ are the same for two networks, then also  $\bar v$ must be the same.
\begin{defn}
	Let the network model $M^{(i)}$ correspond to open-loop transfer $T_{wr}^{(i)}$ and noise spectrum $\Phi_{\bar v}^{(i)}$ for $i=\{1,2\}$.
	Network models $M^{(1)}$ and $M^{(2)}$ are said to be equivalent if
	\begin{equation}
		T_{wr}^{(1)}=T_{wr}^{(2)} \quad \text{and} \quad \Phi_{\bar v}^{(1)}=\Phi_{\bar v}^{(2)}.
	\end{equation}	
	\hfill $\Box$
\end{defn}
In the above definition, $T_{wr}^{(i)}$ and $\Phi_{\bar v}^{(i)}$ are associated with $w$ and $r$ for $i=\{1,2\}$.
There is an implicit assumption in the definition that the $w$ and $r$ are the same for both $i=\{1,2\}$.

The full freedom that is available for transformation of a network model to an equivalent network model is characterized by operations applied to the network equation.
These transformations can be represented by pre-multiplying with a matrix.
Consider a square rational transfer function matrix $P$, then pre-multiplication of network equation \eqref{eq.dgsMatrix} results in
\begin{equation} \label{eq:premulti}
 	P(q) w(t) = P(q) \Big ( G(q) w(t) + u(t) + v(t) \Big ).
\end{equation}
The above pre-multiplication typically leads to a left-hand side unequal to $w(t)$, in which case we need to move terms to the right-hand side until we have $w(t)$ on the left-hand side, i.e.
\begin{equation}
 	w(t) \! = \! (I -P(q)) w(t)  +  P(q) \Big (  G(q) w(t) + u(t) + v(t) \Big ),
\end{equation}
which is denoted as
\begin{equation}
 	w(t) =  G^{(2)}(q) w(t) + u^{(2)}(t) +  v^{(2)}(t)
\end{equation}
where
\begin{equation} \label{eq:transfo}
 	 G^{(2)} = I-P(I-G), \quad  u^{(2)} = Pu, \quad  v^{(2)} = Pv.
\end{equation}
The transfer function matrix $R$ is then transformed as
\begin{equation}\label{eq:transfoR}
 	R^{(2)} = P R.
\end{equation}
A transformation of the noise model is defined in the following way.
When we describe the noise model as $v = H e$, a pre-multiplication with $P$ does not necessarily lead to a \red{proper, monic, stable and stably invertible} filter $P H$.
For that reason  $H^{(2)}$ and $\Lambda^{(2)}$ are obtained through spectral factorization of the transformed noise spectrum
\begin{equation} \label{eq:transfoH}
 	P(e^{i \omega})  \Phi_v(\omega )  P^T(e^{-i \omega} )   =    H^{(2)}(e^{i \omega})  \Lambda^{(2)}   (H^{(2)}(e^{-i \omega}))^T.
\end{equation}
A transformation $P$ that leads to  an appropriate network representation must satisfy some conditions.
\begin{prop} \label{prop:transP}
	Let network model $M^{(1)}$ satisfy Definition \ref{def:netw_model}.
	The transformation $P$ operating on $M^{(1)}$ as defined in \eqref{eq:transfo}, \eqref{eq:transfoR}, and \eqref{eq:transfoH} leads to a network model $M^{(2)}$  that satisfies Definition \ref{def:netw_model} and that is equivalent to $M^{(1)}$ if and only if:
	\begin{enumerate}
		\item $P$ is full rank, and
		\item diag$\left( I-P(I-G^{(1)}) \right)=0$.
	\end{enumerate}
\end{prop}
\textbf{Proof:} Collected in the appendix.
\hfill $\Box$

An interesting feature of the  network transformations is that the response from external variables and process noises to internal variables remains the same.
A pre-multiplication $P$ as defined above leaves the transfer function matrix $T_{wr}$ invariant, since
\begin{equation}
 	T_{wr} = (P(I-G))^{-1} P R,
\end{equation}
where the identity $P^{-1}P=I$ is used.

A pre-multiplication $P$ that leads to a non-hollow $G^{(2)}$ can be used to transform a network.
However, in that situation additional manipulations would be necessary to arrive at a hollow representation. Without loss of generality we will restrict to transformations $P$ that immediately result in a network representation that satisfies  \eqref{def:netw_model}.
There are some restrictions on $P$, but a large freedom in the choice of transformation $P$ is left.
\begin{prop} \label{prop:freeG}
The equivalence transformation presented in \eqref{eq:transfo}, \eqref{eq:transfoR}, and \eqref{eq:transfoH} can transform a network model $M^{(1)}$ with corresponding modules $G^{(1)}$ into a network model $M^{(2)}$ with corresponding modules $G^{(2)}$  using the transformation
\begin{equation}
 	P = (I- G^{(2)})(I-G^{(1)})^{-1}.
\end{equation}
\end{prop}
\textbf{Proof:} Collected in the appendix.
\hfill $\Box$

The proposition allows for $G^{(1)}$ to be transformed into an arbitrary $G^{(2)}$ as long as it is part of a valid network description.
The consequence of transforming  $G^{(1)}$ to an arbitrary $G^{(2)}$ is that the corresponding $R^{(2)}$ will have a  complex structure
\begin{equation} \label{eq:Rtrans}
 R^{(2)} = P R^{(1)} = (I- G^{(2)})(I-G^{(1)})^{-1} R^{(1)}.
\end{equation}
The implication is that when  $G^{(1)}$ is transformed, $ R^{(2)}$ will compensate the changes to keep the node behavior invariant.
This also holds for the noise model, which will contain additional correlations.
Without any further restrictions on the choice of $R$ and $H$,
the modules represented in $G$ contain no information on the dynamic network.
It is the combination of $G,R,H$ that determines the dynamic network.

\subsection{Abstraction} \label{sect:abstraction}
The next step is to extend network equivalence with the option to remove nodes from the representation.
To this end the concept of \emph{network abstraction} is defined next.
This definition is related to the notions of abstraction in \cite{pappas1996towards,woodbury2017well}.
\begin{defn}
	Let network model $M^{(1)}$ be associated with nodes $w^{(1)} \in \mathbb{R}^{L_1}$, external variables $r \in \mathbb{R}^K$, open-loop transfer $T_{wr}^{(1)} \in \mathbb{R}^{L_1 \times K}$, and noise spectrum $\Phi_{\bar v}^{(1)} \in \mathbb{R}^{L_1 \times L_1}$.
	Let network model $M^{(2)}$ be associated with nodes $w^{(2)} \in \mathbb{R}^{L_2}$, external variables $r \in \mathbb{R}^K$, open-loop transfer $T_{wr}^{(2)} \in \mathbb{R}^{L_2 \times K}$, and noise spectrum $\Phi_{\bar v}^{(2)} \in \mathbb{R}^{L_2 \times L_2}$.
	Let $L_2 < L_1$ and let $C$ be the matrix that selects $w^{(2)}$ from $w^{(1)}$, so define $C$ with one 1 per row, zeros everywhere else,  full row rank, and such that $w^{(2)} = Cw^{(1)}$.
	Network model $M^{(2)}$ is said to be an abstraction of $M^{(1)}$ if
	\begin{equation} \label{eq:defabs}
	 	T_{wr}^{(2)} = C T_{wr}^{(1)}, \quad  \Phi_{\bar v}^{(2)} = C \Phi_{\bar v}^{(1)} C^T.
	\end{equation}
	The nodes that are in $w^{(1)}$, but not in $w^{(2)}$ are said to be abstracted from the network.
	\hfill $\Box$
\end{defn}
Constructing an abstraction of a network implies that some nodes are removed from the network representation, while the remaining nodes stay invariant, in the sense that for the same external signal $r$, the second order statistical properties of the remaining node signals are invariant.

The next step is to determine how to obtain an abstraction of a network.
In certain cases, abstracting nodes $\bar w$ from a network can be done by simply pre-multiplying the network equation \eqref{eq.dgsMatrix} with the selection matrix $C$, i.e.
\begin{equation} \label{eq:premulti2}
 	C w(t) = C  \Big ( G(q) w(t) + R(q) r(t) + v(t) \Big ).
\end{equation}
However, this only is an abstraction if the abstracted nodes $\bar w$ no longer appear on the right-hand side of the equation.
If $\bar w$ appears on the right-hand side of \eqref{eq:premulti2} then the abstracted nodes have an influence on the behavior of the nodes in $Cw$, such that \eqref{eq:defabs} cannot hold.
It has to be determined how to define a transformation $P$ such that an abstraction can be obtained by selecting rows from the equation, as in \eqref{eq:premulti2}.

A node $w_i$ influences other nodes through its out-neighbors, and these corresponding modules are located in a column in $G$.
If a node has no influence on the rest of the network, then it has no out-neighbors, and the corresponding column is 0.
Abstracting node $w_i$ requires us to transform the network such that a 0-column is formed by transformation, after which the node can be removed.
By Proposition \ref{prop:freeG} we know that such a transformation always exists.
The abstraction satisfies the relations
\begin{equation} \label{eq:abstra}
 	 G^{(2)} = C  \left ( I-P(I-G) \right ) C^T , \quad  R^{(2)} = CPR.
\end{equation}
A noise model constructed as $CPH^{(1)}$ is a non-square matrix, which is difficult to handle in an identification setting.
Therefore the transformed noise model $H^{(2)}$, $\Lambda^{(2)}$ will be obtained through spectral factorization
\begin{equation} \begin{split}
 	C P(e^{-j \omega})  \Phi_v(\omega )  P^T(e^{j\omega} ) C^T  = \\  H^{(2)}(e^{-j\omega})  \Lambda^{(2)}   (H^{(2)}(e^{j\omega}))^T.
\end{split}\end{equation}
\subsection{Discussion on Identifiability} \label{sec:disc_ident}
In Proposition \ref{prop:freeG} we have seen that $G^{(1)}$ can be transformed into an arbitrary $G^{(2)}$ as long as it is part of a valid network description. This may give rise to the question whether we are not dealing with an unnecessarily overparameterized situation, involving $G$, $R$ and $H$ to describe a dynamic network. However we particularly include the situation that measured external excitation signals enter into physical subsystems of a network, and thus our modules in $G$ can have an intrinsic interpretation in the physical world. The price of abstracting nodes in a network
%
is that $H$ and $R$ can become complex, which may be impossible to identify.
In the identifiability analysis provided in \cite{weerts2018identifiability} the following necessary condition for network identifiability can be found: For a network model set $\{G(\theta),R(\theta),H(\theta), \theta \in \Theta\}$ to be network identifiable it is necessary that at least $L$ entries on each row of $[G(q,\theta) \; R(q,\theta) \; H(q,\theta)]$ are fixed and non-parameterized.
If only the topology is known, then the absent links in the network reflected by zeros in the matrices $G,R,H$ are the only known entries.
In case  $H$ and $R$ become too complex, then the number of zeros is too small for the model to be embedded in a network identifiable model set.
This implies that it is impossible to find a unique estimate for the model structure on the basis of the data.
The approach in the next section is to define a particular abstraction method,  that  leads to abstracted networks that can be embedded in network identifiable model sets.
\section{Abstraction of networks} \label{sect:abstract}
We will now formulate and analyze an abstraction algorithm for dynamic networks that generalizes the  procedure of immersion \cite{dankers2016identification} and the indirect inputs method \cite{linder2017identification}. It starts by dividing the network nodes into a set of nodes $w_\Sc$ that are retained after abstraction and a set of nodes $w_\Zc$ that will be removed. The abstracted network will then allow us to analyze the properties of estimated models when the retained/measured node signals are employed in an identification procedure.
\subsection{Generalized algorithm}
It appears that the action of abstracting nodes in a network is not unique. This is particularly due to the degrees of freedom that exist in transforming network representations to equivalent forms, by premultiplying the system's equations by appropriate transformation matrices. In order to incorporate this freedom in the abstraction, we decompose each set of node signals $\Sc$ and $\Zc$ into two disjunct parts:
\begin{eqnarray}
   \Sc & = & \Lc \cup \tilde\Sc \\
   \Zc & = & \Vc \cup \tilde\Zc.
\end{eqnarray}
The node signals $w_{\tilde \Zc}$ will be abstracted directly by substituting the equation for $w_{\tilde \Zc}$ into the equations for the other node signals. The node signals $w_\Vc$ have the property that they can be \emph{indirectly observed} by the node signals $w_\Lc$, and therefore they can be eliminated from the network by utilizing the equation for nodes $w_\Lc$. The notion of \emph{indirect observation} will be specified after the next step.

Based on these sets, the network can be represented by
\begin{equation} \label{eq:netw_4group}
\begin{bmatrix}
\! w_{\tilde \Sc} \! \\ \! w_\Lc \! \\ \! w_\Vc \! \\ \! w_{\tilde \Zc} \!
\end{bmatrix}
\! = \!
\begin{bmatrix}
\! G_{\tilde \Sc \tilde \Sc} & G_{\tilde \Sc \Lc} & G_{\tilde \Sc \Vc} & G_{\tilde \Sc \tilde \Zc} \! \\
\! G_{\Lc \tilde \Sc} & G_{\Lc \Lc} & G_{\Lc \Vc} & G_{\Lc \tilde \Zc} \! \\
\! G_{\Vc \tilde \Sc} & G_{\Vc \Lc} & G_{\Vc \Vc} & G_{\Vc \tilde \Zc} \! \\
\! G_{\tilde \Zc \tilde \Sc} & G_{\tilde \Zc \Lc} & G_{\tilde \Zc \Vc} & G_{\tilde \Zc \tilde \Zc} \!
\end{bmatrix} \! \! \!
\begin{bmatrix}
\! w_{\tilde \Sc} \! \\ \! w_\Lc \! \\ \! w_\Vc \! \\ \! w_{\tilde \Zc} \!
\end{bmatrix}
\! + \!
\begin{bmatrix}
\! u_{\tilde \Sc} \! \\ \! u_\Lc \! \\ \! u_\Vc \! \\ \! u_{\tilde \Zc} \!
\end{bmatrix}
\! + \!
\begin{bmatrix}
\! v_{\tilde \Sc} \! \\ \! v_\Lc \! \\ \! v_\Vc \! \\ \! v_{\tilde \Zc} \!
\end{bmatrix} \! ,
\end{equation}
where $w_{\tilde \Zc}$ and $w_{\tilde \Sc}$ are defined according to $\tilde \Sc = \Sc \backslash \Lc$ and $\tilde\Zc =  \Zc \backslash \Vc$.

The node signals $w_\Lc$ and $w_\Vc$ are chosen in such a way that the signals $w_\Lc$ serve as indirect observations of the node signals $w_\Vc$, meaning that the signals $w_\Lc$ contain sufficient information of the indirectly observed  signals $w_\Vc$, so as to replace them in an elimination procedure. This is formulated in the following definition.
\begin{defn}[indirect observations] \label{def1}
The node signals $w_\Lc$ serve as indirect observations of the node signals $w_\Vc$ if the transfer function $G_{\Lc \Vc} + G_{\Lc \tilde\Zc}(I-G_{\tilde\Zc \tilde \Zc})^{-1} G_{\tilde \Zc \Vc} $ has full column rank.
\hfill $\Box$
\end{defn}
The property is satisfied if in the network there exists a sufficient number of paths from nodes $w_\Vc$ to nodes $w_\Lc$ that run through nonmeasured nodes only.
An illustration of indirect observations is provided in Section \ref{sec:interp}.
In the remainder of this paper it will be assumed that $w_\Lc$ and $w_\Vc$ are selected to satisfy the full column rank property of the above definition.
Note that this assumption is not restrictive because $\Lc$ and $\Vc$ are chosen by the user such that the assumption is satisfied.
We can always choose $\Lc$ and $\Vc$ as empty sets, so the assumption does not prevent us from constructing an abstracted network.
However, the use of non-empty sets $\Lc$ and $\Vc$ will allow us to use more degrees of freedom in constructing abstracted networks.

Now we can formulate the generalized abstraction algorithm, as follows.
\begin{alg} \label{alg:gen}
Consider a network representation as in \eqref{eq:netw_4group}. Then the following algorithm leads to a network in which nodes $w_\Zc$ are abstracted and nodes $w_\Sc$ are retained:
\begin{enumerate}
\item[a] Solve the fourth equation of \eqref{eq:netw_4group} for $w_{\tilde \Zc}$, and then substitute the result into the other equations, and remove the fourth equation from the network.
\item[b] Solve the second equation of \eqref{eq:netw_4group} for $w_\Vc$, and substitute the result into the first equation.
\item[c] Solve the third equation of \eqref{eq:netw_4group} for $w_\Vc$, and substitute the result into the second equation, and remove the third equation from the network.
\item[d] Remove possible self-loops in the resulting network representation by shifting self-loop terms to the left hand side of the equations and scaling the equations such that an identity matrix remains at the left hand side.
\hfill $\Box$
\end{enumerate}
\end{alg}
The algorithm shows that the essential difference between the nodes in $w_{\tilde \Sc}$ and $w_\Lc $ is how $w_\Vc$ is removed from their equation.
After application of the algorithm, the abstracted network will be represented by
\begin{equation} \label{eq:netw_2group}
\begin{bmatrix}
\! w_{\tilde \Sc} \! \\ \! w_\Lc \end{bmatrix}
\! = \!
\begin{bmatrix}
\! \check G_{\tilde \Sc \tilde \Sc} & \check G_{\tilde \Sc \Lc}  \\
\! \check G_{\Lc \tilde \Sc} & \check G_{\Lc \Lc} \end{bmatrix} \! \! \!
\begin{bmatrix}
\! w_{\tilde \Sc} \! \\ \! w_\Lc
\end{bmatrix}
\! + \!
\begin{bmatrix}
\! \check u_{\tilde \Sc} \! \\ \! \check u_\Lc
\end{bmatrix}
\! + \!
\begin{bmatrix}
\! \check v_{\tilde \Sc} \! \\ \! \check v_\Lc
\end{bmatrix} \! .
\end{equation}
Note that the particular type and ordering of variable substitution in Algorithm \ref{alg:gen} essentially influences the result. This will be illustrated with examples in Section \ref{sec:exam}.
\subsection{Specification through transformations} \label{sect:trans}
Algorithm \ref{alg:gen} can be specified by denoting the algebraic manipulations that generate the substitution and elimination operations in the different steps of the algorithm. We will first describe the substitution operations on the set of equations, and at the end of the procedure address the removal of equations.

In step {\it (a)} the elimination of $w_\Zc$ is performed, which corresponds to applying the transformation matrix
\begin{equation}
 	P^{(1)} = \begin{bmatrix} I&0&0&G_{\tilde \Sc \tilde \Zc}(I-G_{\tilde \Zc \tilde \Zc})^{-1}\\0&I&0&G_{\Lc \tilde \Zc}(I-G_{\tilde \Zc \tilde \Zc})^{-1}\\0&0&I&G_{\Vc \tilde \Zc}(I-G_{\tilde \Zc \tilde \Zc})^{-1}\\0&0&0&(I-G_{\tilde \Zc \tilde \Zc})^{-1} \end{bmatrix} ,
\end{equation}
to the network representation \eqref{eq:netw_4group}. This leads to a new $G$-matrix given by
\begin{equation}
 	G^{(1)} = I-P^{(1)}(I-G),
\end{equation}
where $G$ is partitioned as defined in \eqref{eq:netw_4group}, and $G^{(1)}$ is partitioned in the same way.

In steps {\it (b)-(c)}, we first obtain a new expression for $w_\Vc$ by reverting the expression for $w_\Lc$, and we substitute the original expression for $w_\Vc$ into the expressions for $w_\Lc$.
This corresponds to applying the transformation matrix
\begin{equation}
 	P^{(2)} = \begin{bmatrix}
 		I&0&0&0\\
 		0&I&G_{\Lc\Vc}^{(1)}(I-G_{\Vc\Vc}^{(1)})^{-1}&0\\
 		0&(G_{\Lc\Vc}^{(1)})^\dagger&0&0\\
 		0&0&0&I
 	\end{bmatrix},
\end{equation}
with $(G_{\Lc\Vc}^{(1)}) = G_{\Lc \Vc} + G_{\Lc \tilde\Zc}(I-G_{\tilde\Zc \tilde \Zc})^{-1} G_{\tilde \Zc \Vc} $, and $(G_{\Lc\Vc}^{(1)})^\dagger$ denoting its left-inverse,
and $(G_{\Vc \Vc}^{(1)}) = G_{\Vc \Vc} + G_{\Vc \tilde\Zc}(I-G_{\tilde\Zc \tilde \Zc})^{-1} G_{\tilde \Zc \Vc} $,
 leading to
\begin{equation}
 	G^{(2)} = I-P^{(2)}(I-G^{(1)}),
\end{equation}
where  $G^{(2)}$ has the same  partitioning as $G^{(1)}$. Note that the left-inverse exists due to the indirect observations property of $w_\Lc$, as formulated in Definition \ref{def1}.

The remaining part of steps {\it (b)-(c)} is now to substitute the new expression for $w_\Vc$ in the first equation for $w_\Sc$, thereby eliminating the dependency of this expression on $w_\Vc$.
This is achieved by applying the transformation matrix
\begin{equation}
 	P^{(3)} = \begin{bmatrix}
 		I&0&G^{(2)}_{\tilde \Sc \Vc}&0\\
 		0&I&0&0\\
 		0&0&I&0\\
 		0&0&G^{(2)}_{\tilde \Zc \Vc}&I
 	\end{bmatrix},
\end{equation}
such that
\begin{equation}
 	G^{(3)} = I-P^{(3)}(I-G^{(2)}),
\end{equation}
where $G^{(3)}$ has the same partitioning as $G^{(2)}$.
The additional term $G^{(2)}_{\tilde \Zc \Vc}$ that is added in the fourth row of $P^{(3)}$
ensures that in the transformed network all columns that correspond to $w_\Zc$ are zero in $G^{(4)}$, including for the equations that will be removed.

Step {\it (d)} of the Algorithm is addressed by removing self-loops in the resulting network representation, by applying a diagonal transformation matrix $P^{(4)}$ with diagonal elements \begin{equation} \label{eq:P4}
 	P^{(4)}_{jj} = \frac{1}{1-G^{(3)}_{jj}}
\end{equation}
and being $0$ elsewhere.

The total transformation that is applied to the network representation is now given by
\begin{equation}
 	P^{(abs)} = P^{(4)}P^{(3)}P^{(2)}P^{(1)},
\end{equation}
which leads to a $G$-matrix of the transformed network representation that is structured according to
\begin{equation} \label{eq:G4}
 G^{(4)} := (I-P^{(abs)}(I-G)) =
 \begin{bmatrix}
 \check G_{\tilde \Sc \tilde \Sc} & \check G_{\tilde \Sc \Lc} & 0 & 0  \\
 \check G_{\Lc \tilde \Sc} & \check G_{\Lc \Lc} & 0 & 0 \\
 \check G_{\Vc \tilde \Sc} & \check G_{\Vc \Lc} & 0& 0  \\
 \check G_{\tilde \Zc \tilde \Sc} & \check G_{\tilde \Zc \Lc} & 0 & 0
\end{bmatrix}
\end{equation}

The abstracted network now results by selecting the first two block rows and columns in the matrix $G^{(4)}$, thereby removing the equations for the unmeasured/abstracted node variables $w_\Vc$ and $w_{\tZc}$.

\begin{prop} \label{prop:eq_alg_trans}
When applying the abstraction procedure of Algorithm \ref{alg:gen} to a dynamic network given by (\ref{eq:netw_4group}), the obtained abstracted network is the same as the abstracted network given by
(\ref{eq:netw_2group}) with
\begin{align*}
	\begin{bmatrix}
		\check G_{\tilde \Sc \tilde \Sc} & \check G_{\tilde \Sc \Lc}  \\
		\check G_{\Lc \tilde \Sc} & \check G_{\Lc \Lc}
	\end{bmatrix}
	& =
	\begin{bmatrix}	I&0	& 0 & 0 \\ 0 & I & 0 & 0 \end{bmatrix}
	(I-P^{(abs)}(I-G))
	\begin{bmatrix}	I&0	\\ 0 & I \\ 0 & 0 \\ 0 & 0 \end{bmatrix}
	\\
	\begin{bmatrix}
		\check u_{\tilde \Sc} \\ \check u_\Lc
	\end{bmatrix}
	& =
	\begin{bmatrix}	I&0	& 0 & 0 \\ 0 & I & 0 & 0 \end{bmatrix} P^{(abs)}
	\begin{bmatrix}
		 u_{\tSc} \\  u_\Lc \\  u_\Vc \\  u_{\tZc}
	\end{bmatrix}
	\\
	\begin{bmatrix}
		\check v_{\tilde \Sc} \\ \check v_\Lc
	\end{bmatrix}
	& =
	\begin{bmatrix}	I&0	& 0 & 0 \\ 0 & I & 0 & 0 \end{bmatrix} P^{(abs)}
	\begin{bmatrix}
		 v_{\tSc} \\  v_\Lc \\  v_\Vc \\  v_{\tZc}
	\end{bmatrix} .
\end{align*}
\end{prop}
\textbf{Proof:} Collected in the appendix.
\hfill $\Box$

\subsection{Interpretations and discussion}
\label{sec:interp}
\begin{figure}[htb]
	\centering
	\includegraphics[width=0.5\columnwidth]{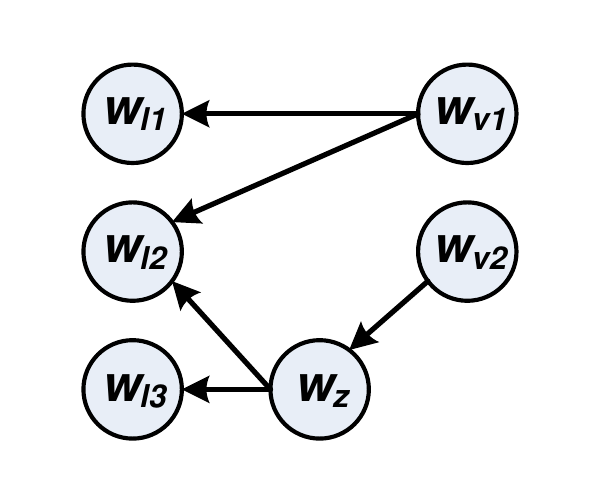}
	\caption{Example network with $\Vc=\{v_1,v_2\}$, $\tilde \Zc=\{z\}$, $\Lc=\{l_1,l_2,l_3\}$, where the full rank condition is satisfied.}
	\label{fig:indinp_show}
\end{figure}

Compared to selecting a set of nodes $w_\Zc$,
the particular choice of the sets of nodes $w_\Lc$ and $w_\Vc$ creates additional degrees of freedom in the problem of constructing an abstracted network, in which the nodes $w_\Zc$ (including $w_\Vc$) are removed.
The mechanism that is used is that the network equation for the node signals $w_\Lc$ is inverted to become an equation that describes the node signals $w_\Vc$.
This equation is then subsequently used to substitute and eliminate the $w_\Vc$ signals from the abstracted network.
In order to be able to use the network equation for $w_\Lc$ in this way, it needs to capture full information on the node signals $w_\Vc$.
This is reflected in the property of \emph{indirect observations}, and the required full column rank property of $G_{\Lc \Vc} + G_{\Lc \tilde\Zc}(I-G_{\tilde\Zc \tilde \Zc})^{-1} G_{\tilde \Zc \Vc} $, as formulated in Definition \ref{def1}.
This full rank property implies that $dim(w_\Lc) \geq dim(w_\Vc)$.
It is generically satisfied if there are dim$(w_\Vc)$ vertex-disjoint paths present from  $w_\Vc$ to  $w_\Lc$ that run through nonmeasured/abstracted nodes only \citep{Woude1991,Hendrickx&Gevers&Bazanella_TAC:19}.
An example of the full rank assumption is shown in Figure \ref{fig:indinp_show}.
In the figure there are the two vertex-disjoint paths $w_{v1} \rightarrow w_{l1}$ and $w_{v2} \rightarrow w_z \rightarrow w_{l3}$ for two nodes that are indirectly observed.
In this case actually any selection of two nodes from $\{w_{l_1}, w_{l_2}, w_{l_3}\}$ would be sufficient to act as indirect observations of $\{w_{v_1}, w_{v_2}\}$.

The network abstraction introduced here, generalizes two earlier introduced abstraction algorithms. For the particular choice, $\Lc= \emptyset$ and $\Vc = \emptyset$, Algorithm \ref{alg:gen} describes the method of network immersion, as introduced in \cite{dankers2016identification}, and developed for the situation $R=I$. In that case steps {\it (2)-(3)} of the algorithm become obsolete.

If $w_\Lc$ is restricted to consist of nodes that are out-neighbors of $w_\Vc$,
and $w_\Vc$ does not contain in-neighbors $of w_{\tilde{\mathcal{Z}}}$,
and $G_{\Lc \Vc}$ has full column rank, then Algorithm \ref{alg:gen} describes the indirect inputs method as defined in \cite{linder2017ifac}, which has been developed for the situation $R=I$.
If $\Lc=\emptyset$, then the indirect inputs method is equivalent to the immersion method.

Because of well-posedness of the original network, all terms in the transformation matrices $P^{(1)}\cdots P^{(4)}$ are proper, except possibly for the term $(G_{\Lc\Vc}^{(1)})^\dagger$ which may be non-proper. This implies that the introduction of the sets $\Lc$ and $\Vc$ may lead to a final abstracted network representation that is non-proper. Properness of the resulting network representation is guaranteed if $\Lc= \emptyset$.

\subsection{Identifiability analysis}
As discussed in Section \ref{sec:disc_ident}, it is impossible to formulate an identifiable model set for network representations of high complexity.
The underlying objective of the particular abstraction algorithm  is to limit the complexity of the abstracted network.
An evaluation of the structure of network representations obtained with the abstraction algorithm is made.
In this way we can guarantee that an identifiable model set can be defined for the abstracted network.

A sufficient condition for network identifiability that can easily be verified is that every node has an independent external excitation.
This is achieved when the columns of $\check R$ in
\begin{equation} \label{eq:checkR}
 	\check u(t) = \check R(q) r
\end{equation}
can be permuted to arrive at a matrix with a leading diagonal \citep{weerts2018identifiability}.
In order to verify whether this can be achieved we need to evaluate the structure of $\check R$.
The abstracted network generated by Algorithm \ref{alg:gen} corresponds to the transformation $P^{(abs)}$, such that
\begin{equation}
 	\check R = \begin{bmatrix}	I&0	 \end{bmatrix} P^{(abs)} R.
\end{equation}
Since $P^{(abs)}$ and $R$ are formulated in terms of the original network, we can formulate conditions for network identifiability based on the original topology.
This is formally done in the next proposition.

\begin{prop} \label{prop:diagR}
	Consider the abstracted network \eqref{eq:netw_2group} obtained by abstracting the original network \eqref{eq.dgsMatrix} with Algorithm \ref{alg:gen} by using the sets of nodes $\tSc,\Lc,\Vc,\tZc$.
	The representation of the external excitations is $\check u(t) = \check R(q) r$.
	The matrix $\check R(q)$ can be given a leading diagonal by column operations if the original network is such that
	\begin{enumerate}[leftmargin=*]
		\item $R_{\tilde \Sc \tilde \Sc}$ is diagonal, and $r_{\tilde \Sc}$ is not an in-neighbor of nodes other than $w_{\tilde \Sc}$,
		\item $R_{\Vc \Vc}$ is diagonal, and  $r_{\Vc}$ is not an in-neighbor of nodes other than $w_{\Vc}$, and
		\item $G_{\Lc\Vc}$ is diagonal, $G_{\Lc\tZc}=0$, $G_{\Vc\Vc}=0$, and $G_{\Vc\tZc}=0$.
	\end{enumerate}
\end{prop}
\textbf{Proof:} Collected in the appendix.
\hfill $\Box$

The proposition implies that abstracted networks that are obtained by Algorithm \ref{alg:gen} can be embedded in network identifiable model sets, under some restrictions on the original network.
Here we have analyzed network identifiability of all modules in the abstracted network using sufficient conditions.
The result may be extended by using less restrictive conditions that make use of the topology present in $G$ \cite{weerts2018identifiability}.
Moreover, for consistency of the module of interest only network identifiability of that particular module is necessary.
Conditions for network identifiability of a particular module are less restrictive than conditions for network identifiability of all modules \citep{weerts2018_cdc}, which could further reduce the imposed conditions on the structure of the network.

\section{Abstraction applied to an example network} \label{sec:exam}
\begin{figure}[htb]
	\centering
	\includegraphics[width=0.85\columnwidth]{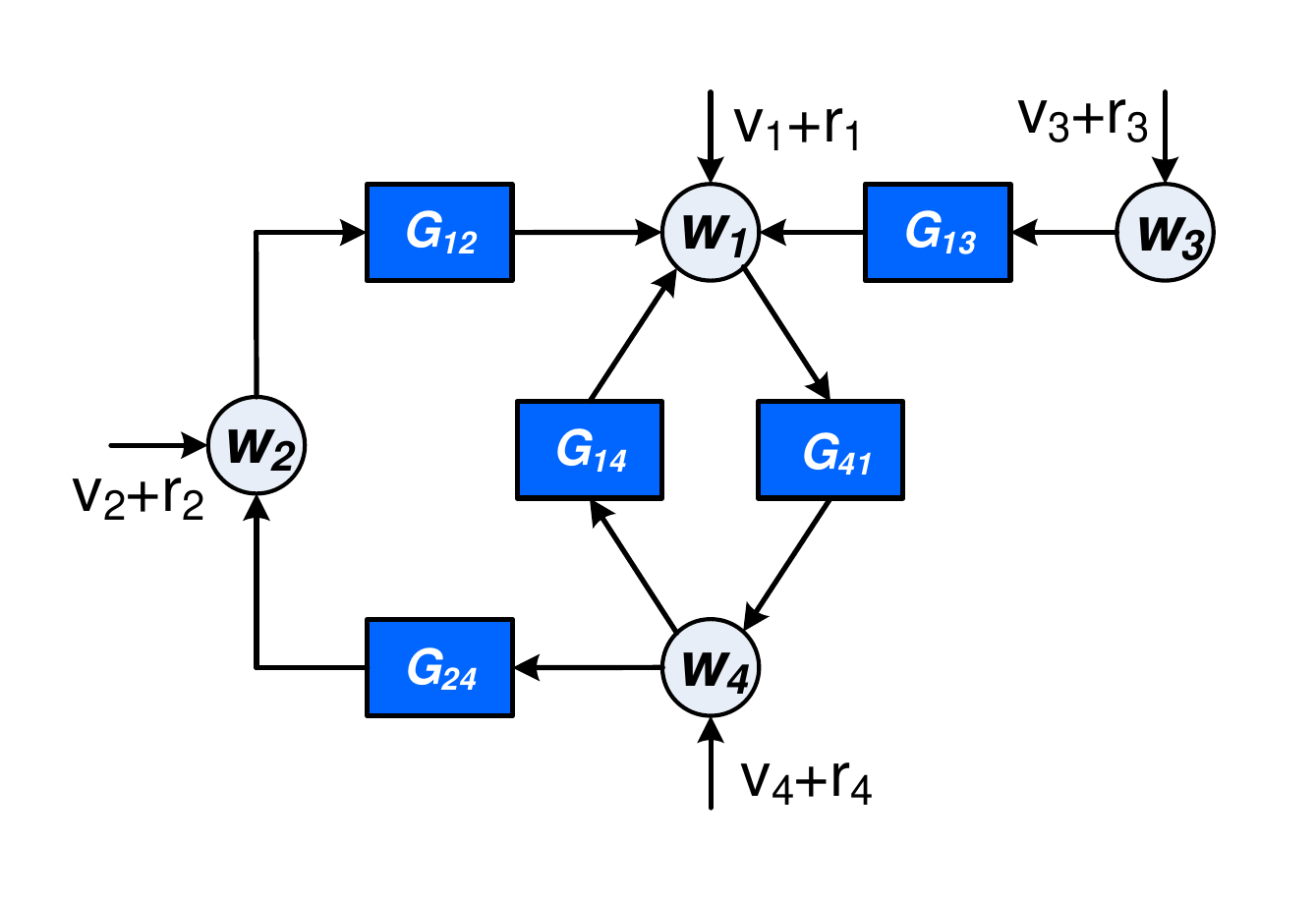}
	\caption{Example network to illustrate abstraction.}
	\label{fig:netw_abs}
\end{figure}

In this section we will provide an example to illustrate some of the options that are available in network abstraction.
Consider the network in Figure \ref{fig:netw_abs} where the nodes are described by the following equations
\begin{eqnarray}
w_1 & = & G_{12}w_2+G_{13}w_3+G_{14}w_4+r_1+v_1 \label{eqx1}\\
w_2 & = & G_{24}w_4+r_2+v_2 \label{eqx2}\\
w_3 & = & r_3+v_3 \label{eqx3}\\
w_4 & = & G_{41}w_1+r_4+v_4 \label{eqx4}
\end{eqnarray}

If we would like to abstract node $w_4$, e.g. because this node signal cannot be measured, then we have different options for doing so.
The set of retained nodes is $\Sc = \{1,2,3\}$ and the set of removed nodes is $\Zc = \{4\}$.

\begin{figure}[htb]
	\centering
	\includegraphics[width=0.85\columnwidth]{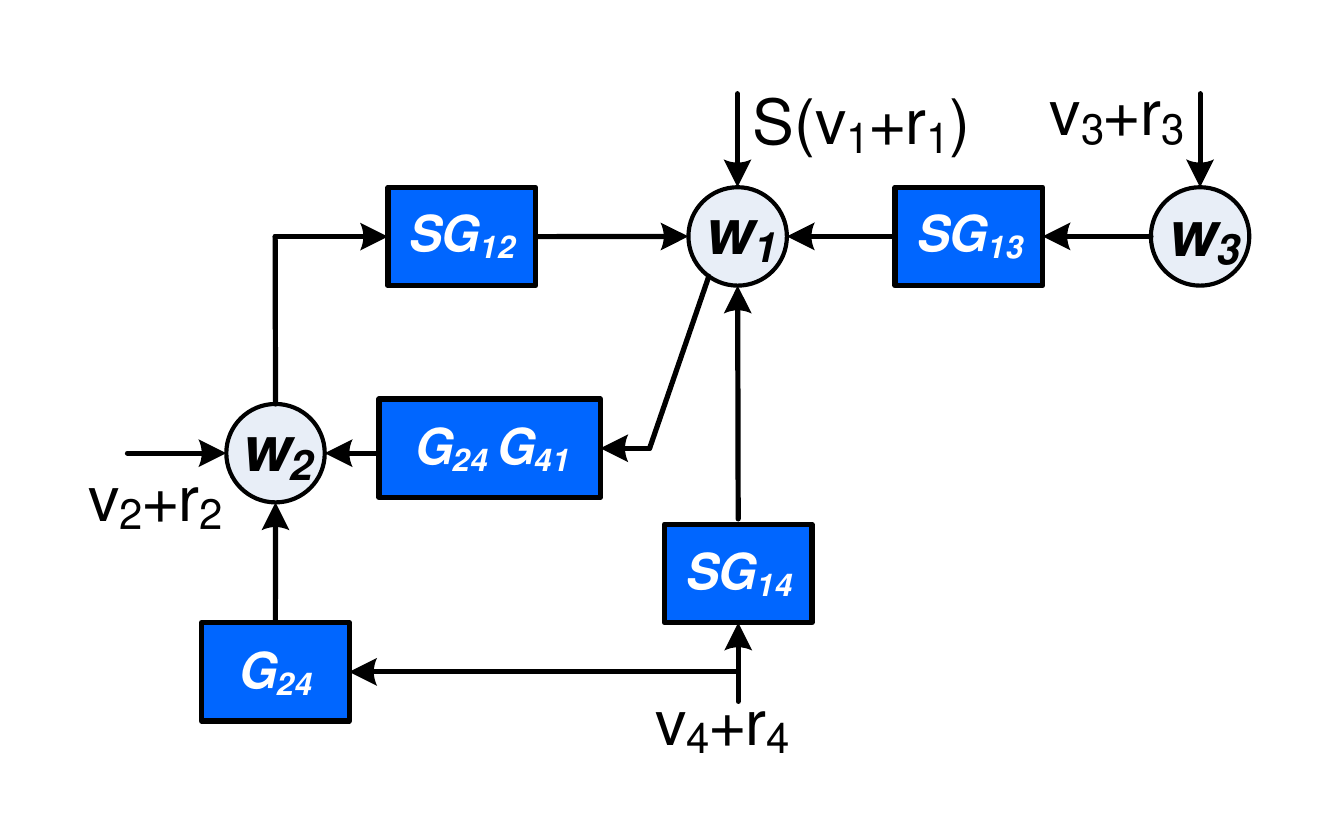}
	\caption{Network obtained after abstracting node $w_4$ through immersion.}
	\label{fig:netw_imm}
\end{figure}

We have to make a choice on whether nodes are used as indirect observations.
If we choose that there are no indirect observations, then $\Lc=\Vc=\emptyset$, such that $\tSc = \{1,2,3\}$ and $\tZc = \{4\}$.
We eliminate $w_4$ from the system equations by substituting its expression \eqref{eqx4} in the expressions of the nodes that are retained \eqref{eqx1}-\eqref{eqx3}.
This leads to a new set of equations given by
\begin{eqnarray*}
w_1 &=& G_{14}G_{41} w_1\! +\! G_{12} w_2\! +\! G_{13} w_3\! +\! r_1\! +\! v_1\! +\! G_{14}(r_4\! +\! v_4) \\
w_2 & = & G_{24}G_{41} w_1 + r_2+v_2 + G_{24}(r_4+v_4) \\
w_3 & = & r_3+v_3
\end{eqnarray*}
which induces a self-loop around $w_1$. This can be compensated for by moving the $w_1$-dependent term to the left hand side, and rewriting the equation for $w_1$ as
\[ w_1 = S[G_{12} w_2 + G_{13} w_3 + r_1 + v_1 + G_{14}(r_4+v_4)] \]
with $S:=(1-G_{14}G_{41})^{-1}$. As a result the abstracted network is obtained and sketched in Figure \ref{fig:netw_imm}.
This way of eliminating node $w_4$ is referred to as \emph{immersion} \citep{dankers2016identification}, and comes down to lifting each path in the original network that contains the node signal that is eliminated.
After removing the abstracted node signals, the remaining node signals are invariant.

\begin{figure}[htb]
	\centering
	\includegraphics[width=0.85\columnwidth]{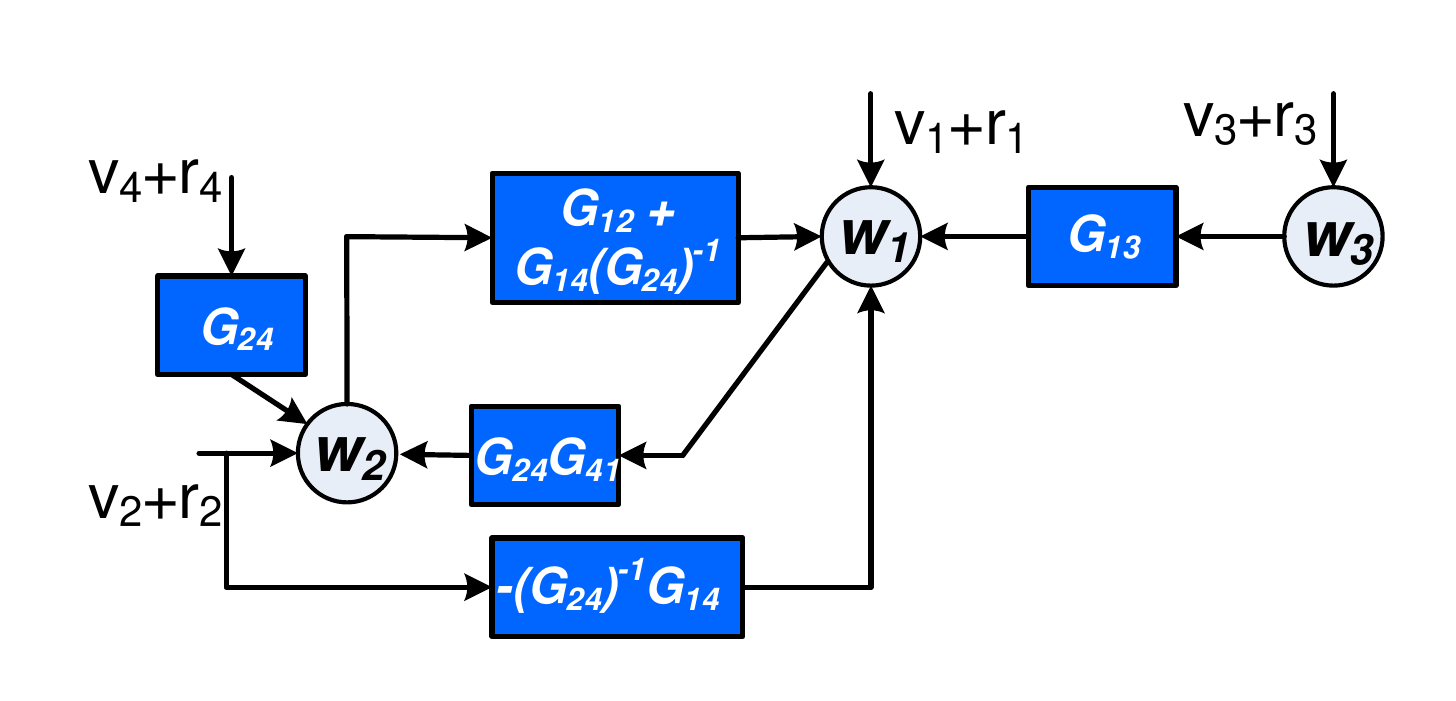}
	\caption{Modification of the network depicted in Figure \ref{fig:netw_abs}, obtained after removal of node $w_4$ by the indirect inputs method.}
	\label{fig:netw_ind}
\end{figure}

As an alternative for the chosen abstraction, we can choose $w_2$ as an indirect observation of $w_4$, such that $\Lc=\{2\}$, $\Vc=\{4\}$, $\tSc=\{1,3\}$, and $\tZc = \emptyset$.
In this situation, node signal $w_4$ is abstracted by utilizing the expression for $w_2$.
We rewrite equation (\ref{eqx2}) as
\[ w_4 = G_{24}^{-1} (w_2 - v_2 -r_2) \]
and substitute this into (\ref{eqx1}) to obtain the expression for $w_1$
\[ w_1 = G_{12} w_2 + G_{13} w_3 + G_{14} \underbrace{G_{24}^{-1}( w_2 - r_2 -v_2)}_{w_4} + r_1 + v_1. \]
In order to obtain the new expression for $w_2$ we directly substitute the expression for $w_4$  (\ref{eqx4}) into the expression for $w_2$ (\ref{eqx2}).
The abstracted network is sketched in Figure \ref{fig:netw_ind}, and given by the following equations:
\begin{align*}
w_1 = & (G_{12} + G_{14}G_{24}^{-1})w_2 + G_{13} w_3 + r_1 + v_1  \\ & - G_{14}G_{24}^{-1}(r_2 + v_2) \\
w_2  = & G_{24}G_{41} w_1 + r_2+v_2 + G_{24}(r_4+v_4) \\
w_3  = & r_3+v_3
\end{align*}
This alternative method of eliminating node variable $w_4$ is referred to as the \emph{indirect inputs method} introduced in \cite{linder2017identification}. The principle idea is that
the out-neighbor of a node that needs to be abstracted contains information about that  node. Then the equation of the out-neighbor is manipulated in order to obtain an explicit expression for the node to be abstracted, which is then used to eliminate the node from the network. A major difference with the method of immersion is that the inverse of modules may appear in the resulting network representation.

It can be observed that the network topology and module dynamics can change when nodes are abstracted from the network.
In particular the module $G_{13}$ has changed to $SG_{13}$ when the immersion method is applied, while it remains invariant when the indirect inputs method is applied.
This is going to be important when considering the problem of identifying a local module on the basis of a restricted set of measured node signals, as will be discussed in the next section.

\section{Identification setting for invariant modules} \label{sect:invariance}
We have introduced an algorithm to perform network abstraction.
The remaining question to answer is how this will help to select nodes for the identification of the module of interest.
A central point in our reasoning will be the invariance of the target module in the abstracted network. Although for consistent identification of the target module it is not strictly necessary to have target module invariance, cf. e.g. the indirect type identification methods of \cite{bazanella2017identifiability,Gevers&etal_sysid:18,Hendrickx&Gevers&Bazanella_TAC:19} or the Wiener-filter based method of
\cite{materassi2015identification,Materassi&Salapaka:19}, invariant target modules are very attractive in two-stage methods \citep{dankers2016identification,linder2017identification}, and they are indispensable in direct methods \citep{dankers2016identification} that have the potential to provide consistent and maximum likelihood (and thus minimum variance) results.
In \cite{dankers2016identification} and \cite{linder2017identification} two different abstraction methods have been used to select node signals to be measured for identification of a target module, based on the invariant module principle. The prime reasoning and the formulation of generalized results are presented next.

\subsection{Direct identification setup}
If the target module to be identified is $G_{ji}(q)$, then a MISO identification setup on the basis of the abstracted network can be formulated in the following way.
Node $w_j$ is used as output, and the following nodes are inputs: $w_{\tSc \setminus j}$ with $i \in \Sc$, $w_{\Lc}$, and possibly additional external excitations.
In this way an identification algorithm can provide us with a consistent estimate of the modules of the abstracted network $\breve G_{jk}$ for all $k \in \Sc$, provided that some regularity conditions are satisfied, among which sufficient excitation properties of the measured signals.
We have seen in the examples of the previous section that a module may remain unchanged after abstraction for particular choices of the sets $\tSc, \Lc, \Vc, \tZc$, i.e.
\begin{equation} \label{eq:Ginv}
 	G_{ji}(q) = \breve G_{ji}(q).
\end{equation}
If the modules of the abstracted network are estimated consistently, and the module of interest has remained invariant in the abstracted network, then the module of interest is estimated consistently.
In the remainder of this section we will address the problem under which conditions the abstracted network has the mentioned invariance property \eqref{eq:Ginv} of $G_{ji}$.

\subsection{Invariance of the module $G_{ji}$}
When applying immersion as a specific abstraction algorithm, it has been analyzed in \cite{dankers2016identification} under which conditions on the set of retained node signals, a particular module in the network will remain invariant.

\begin{prop}[\cite{dankers2016identification}] \label{prop.setD}
Consider a dynamic network as defined in \eqref{eq.dgsMatrix}, and let  $G_{ji}(q)$ be the module of interest.
Denote with $\breve{G}_{ji}(q,\tSc,\Lc,\Vc,\tZc)$ the module $\breve G_{ji}$ in the network that is abstracted using Algorithm \ref{alg:gen} with the sets $\tSc,\Lc,\Vc,\tZc$.
Let the network abstraction be performed through immersion, i.e. with $\Lc=\Vc=\emptyset$.
Define the set $\mathcal{D}_j = \tSc \setminus j$.
Then
\begin{equation}
 	\breve{G}_{ji}(q,\tSc,\Lc,\Vc,\tZc) = G_{ji}(q)
\end{equation}
if $\mathcal{D}_j$ satisfies the following conditions:
\begin{enumerate}
\item $i \in \mathcal{D}_j$, $j \notin \mathcal{D}_j$, \label{cond.ijInD}
\item every path from $w_i$ to $w_j$, excluding the path $G_{ji}$, goes through a node $w_k$, $k \in \mathcal{D}_j$, \label{cond.wiTowj}
\item every loop from $w_j$ to $w_j$ goes through a node $w_k$, $k \in \mathcal{D}_j$. \label{cond.wjTowj}  \hfill $\Box$
\end{enumerate}
\end{prop}

According to this proposition, there are two situations that need to be checked for guaranteeing module invariance: parallel paths and loops around the output. Every path that connects input and output parallel to the target module, and every loop around the output should be ``blocked'' by another node that is retained in the abstracted network.

\subsection{Generalization of invariance conditions}
The following two examples illustrate the parallel paths and loops around the outputs, leading to a generalization of Proposition \ref{prop.setD} that extends the applicability from the immersion abstraction algorithm to the generalized abstraction algorithm.
In the next examples, noise-free networks are used in order to stick to the core reasoning.

\begin{figure}
	\centering
	\includegraphics[width=0.49\columnwidth]{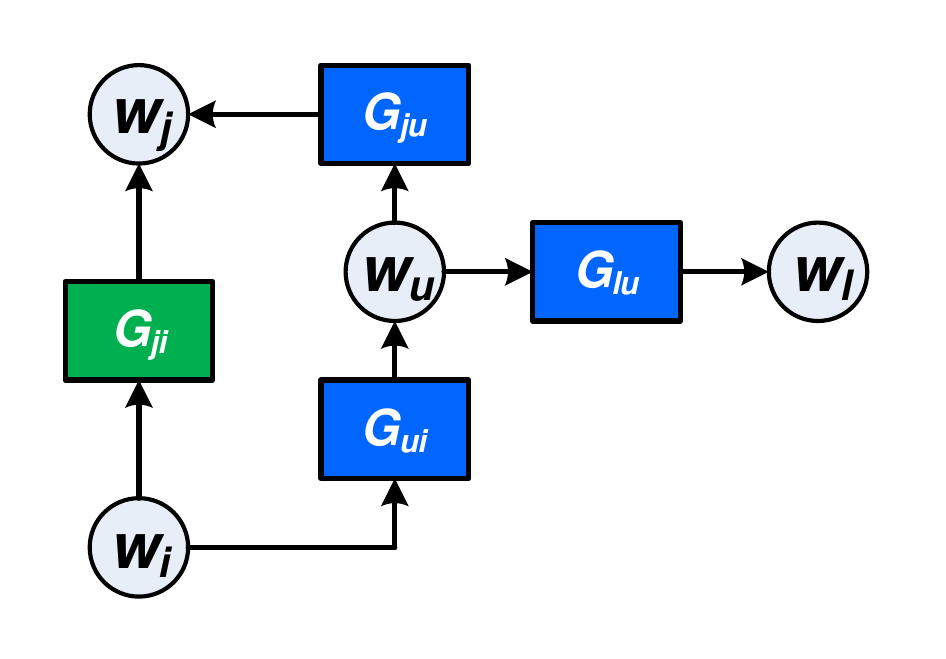}
	\includegraphics[width=0.49\columnwidth]{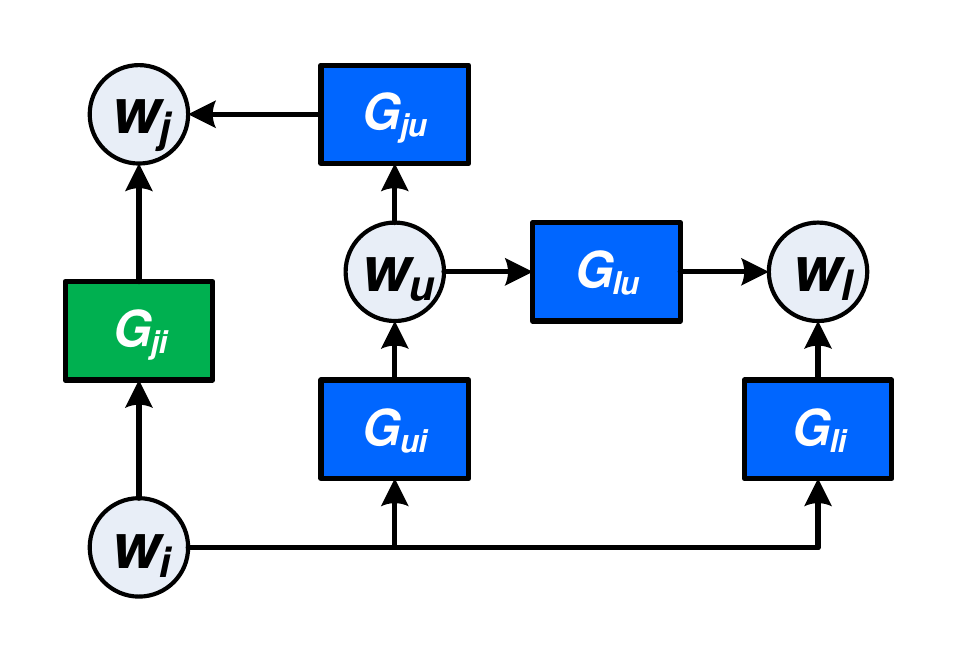}
  \caption{Networks to illustrate issues with parallel paths  when abstracting.}
  \label{fig:ex_imm_ind}
\end{figure}
\begin{exmp}[Parallel paths]\label{ex:ppaths}
Consider the left network in  Figure \ref{fig:ex_imm_ind} and the module of interest $G_{ji}$.
Paths that run in parallel to this module, i.e. paths from $w_i$ to $w_j$, may lead to changes in the module of interest during abstraction.
If $w_u$ is removed using immersion, with $\Lc = \emptyset,\Vc = \emptyset$, then $w_u = G_{ui} w_i$ is substituted into the equation for $w_j$, such that the dynamics of modules $G_{ju}$ and $G_{ui}$ are merged with module of interest $w_i \rightarrow w_j$, i.e.
\begin{equation}
 	w_j = (G_{ji} + G_{ju}G_{ui}) w_i.
\end{equation}
As stated in Proposition \ref{prop.setD}, a way to prevent these parallel paths from changing the module of interest is by including a node in every parallel path in the abstracted network, for example by measuring $w_u$.

An alternative way of removing $w_u$ is to include $w_l$ as an indirect observation of $w_u$, i.e. by choosing $w_\Lc=w_l$, $w_\Vc=w_u$.
In this case the node $w_u$ is substituted with $w_u = G_{lu}^{-1} w_l$ such that
\begin{equation}
 	w_j = G_{ji} w_i + G_{ju} G_{lu}^{-1} w_l.
\end{equation}
The substitution uses an equation that does not contain $w_i$, such that $G_{ji}$ remains invariant.
Apparently, it is not strictly necessary to include a node in every parallel path in the abstracted network.
An indirect observation of a node in the parallel path may be used to \emph{block} this path.

When an additional path $w_i \rightarrow w_l$ exists as in the right network in Figure \ref{fig:ex_imm_ind}, the situation changes.
Now the equation for node $w_l$ depends on $w_i$, and if the unknown node $w_u$ is eliminated using the indirect observation $w_l$, then an additional contribution from $w_i$ appears such that the module of interest is changed, i.e.
\begin{equation}
 	w_j = (G_{ji} -G_{ju}G_{lu}^{-1}G_{li})w_i + G_{ju}G_{lu}^{-1} w_l,
\end{equation}
where $w_u = G_{lu}^{-1} (w_l-G_{li} w_i)$ is used.
If in the left network of Figure \ref{fig:ex_imm_ind} there is no path from $w_u$ to $w_l$, then $w_l$ cannot be used as an indirect observation.
\hfill$\square$
\end{exmp}

From the example it can be observed that the nodes used as indirect observations, i.e. $w_\Lc$, should not have $w_i$ as an in-neighbor.

\begin{figure}
	\centering
	\includegraphics[width=0.49\columnwidth,trim={0.8cm 0 0.7cm 0},clip]{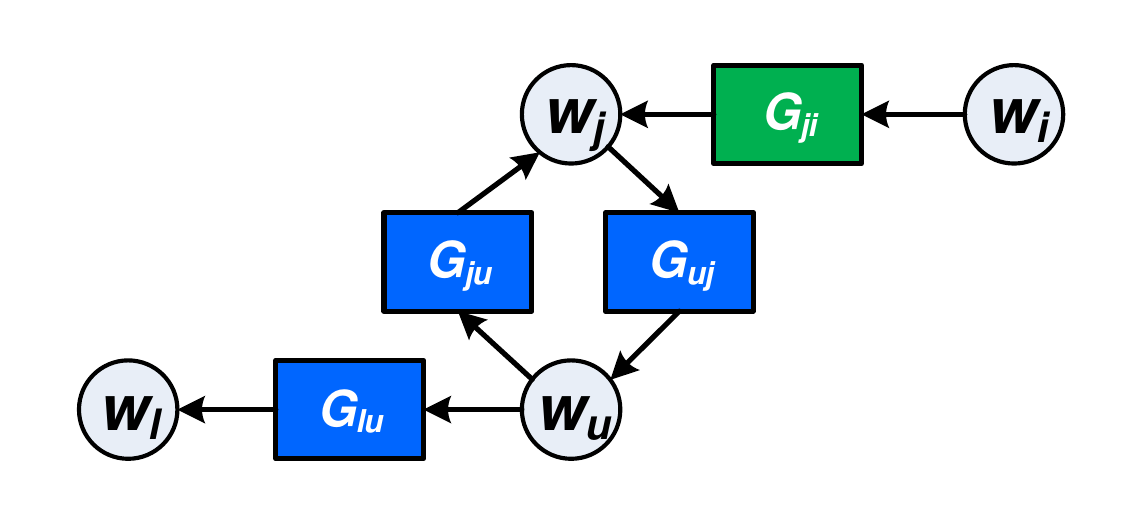}
	\includegraphics[width=0.49\columnwidth,trim={0.8cm 0 0.7cm 0},clip]{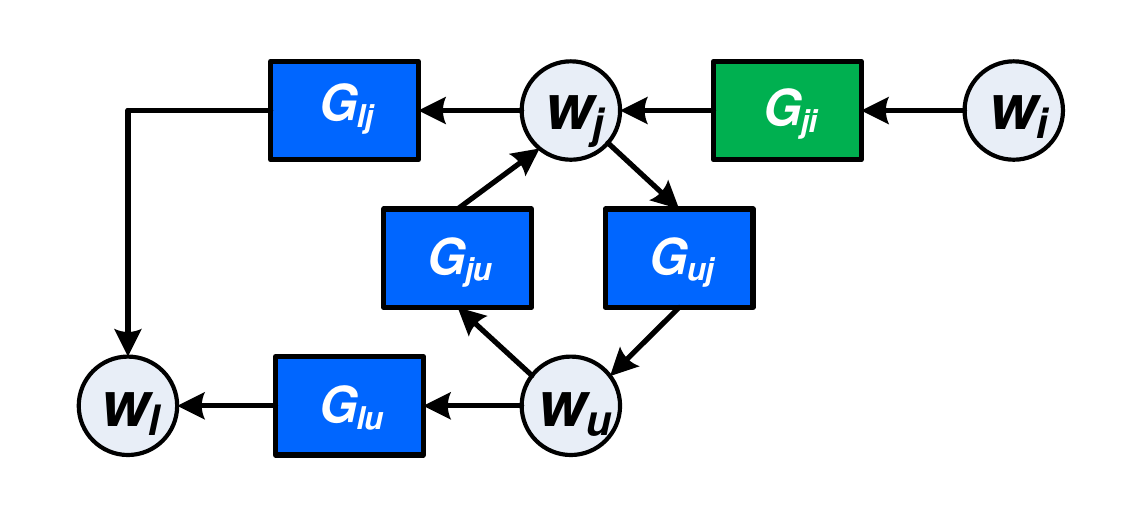}
  \caption{Networks to indicate issues with self-loops when making abstractions.}
  \label{fig:ex_imm_ind_loop}
\end{figure}
\begin{exmp}[Self-loops]\label{ex:selfloops}
Consider the left network in Figure \ref{fig:ex_imm_ind_loop} and suppose the module of interest is $G_{ji}$.
Paths that run as a loop around the output of this module, i.e. paths from $w_j$ to $w_j$, may lead to changes in the module of interest during abstraction.
If the node $w_u$ of the left network in Figure \ref{fig:ex_imm_ind_loop} is eliminated by immersion, using $\Lc=\emptyset$ and $\Vc=\emptyset$, then abstraction leads to the following.
The equation $w_u = G_{uj} w_j$ is substituted into the equation for $w_j$, after which a self-loop around $w_j$ is resolved.
This leads to the following change in the module of interest
\begin{equation}
 	w_j = \frac{G_{ji}}{1-G_{ju}G_{uj}} w_i .
\end{equation}
As stated in Proposition \ref{prop.setD}, a way to prevent these loops around the output from changing the module of interest is by including a node in every such loop around $w_j$ in the abstracted network, for example by measuring $w_u$.

An alternative way of removing $w_u$ is to include $w_l$ as an indirect observation of $w_u$, i.e. by choosing $w_\Lc = w_l$ and $w_\Vc = w_u$.
In this case the $w_u$ is substituted for $w_u = G_{lu}^{-1}w_l$ such that
\begin{equation}
 	w_j = G_{ji} w_i + G_{ju} G_{lu}^{-1} w_l.
\end{equation}
The substitution uses an equation that does not contain $w_j$, such that no self-loop has to be resolved, and $G_{ji}$ remains invariant.
It is thus not strictly necessary to include a node in every loop around $w_j$ in the abstracted network.
An indirect observation of a node in a loop around $w_j$ may be used to \emph{block} this path.

If instead there is a direct link $w_j \rightarrow w_l$ like in the right network of Figure \ref{fig:ex_imm_ind_loop}, then $w_l$ depends directly on $w_j$, and using this equation for elimination of $w_u$ would again lead to a dependence of $w_j$ on itself in the abstracted network, i.e.
\begin{equation}
 	w_j = G_{ji} w_i + G_{ju} G_{lu}^{-1} (w_l - G_{lj} w_j),
\end{equation}
where $w_u$ is substituted for  $w_u = G^{-1}_{lu} (w_l - G_{lj} w_j)$.
The self-loop should be resolved, leading to
\begin{equation}
 	w_j  = \frac{G_{ji}}{1+G_{ju} G_{lu}^{-1}G_{lj}}  w_i + \frac{ G_{ju} G_{lu}^{-1}}{1+G_{ju} G_{lu}^{-1}G_{lj}} w_l
\end{equation}
where it is obvious that the module of interest has changed.
\hfill$\square$
\end{exmp}

In conclusion, for verifying module invariance in abstracted networks obtained by Algorithm \ref{alg:gen} we have to consider the following.
It is not sufficient to only consider parallel paths from $w_i$ to $w_j$ and loops from $w_j$ to $w_j$ that appear in the data generating system.
We have to also consider indirect observations of the nodes that are part of parallel paths and loops around the output.
Paths from $w_i$ and $w_j$ to the indirect observations $w_\Lc$ also have to be considered to avoid merging of paths and to keep $G_{ji}$ invariant under the transformation.
These observations lead to the following formal result.

\begin{thm} \label{prop:cond_gen}
Consider a dynamic network as defined in \eqref{eq.dgsMatrix}, and let  $G_{ji}(q)$ be the module of interest.
Denote with $\breve{G}_{ji}(q,\tSc,\Lc,\Vc,\tZc)$ the module $\breve G_{ji}$ in the abstracted network that is obtained using Algorithm \ref{alg:gen} with the sets $\tSc,\Lc,\Vc,\tZc$.
Assume that nodes $w_\Lc$ act as indirect observations of nodes $w_\Vc$ according to Definition \ref{def1}, and that  $\{i,j\} \subset \tilde \Sc$.
Define the sets $\Jc = \{j\} \cup \Lc$ and $\Kc =  \Vc \cup \tilde \Sc \setminus \{j\}$.
Then
\begin{equation}
	\breve G_{ji}(q,\tSc,\Lc,\Vc,\tZc) = G_{ij}(q) 	
\end{equation}
if the following conditions on the sets $\tilde \Sc$, $\Lc$ and $\Vc$ are satisfied:
\begin{enumerate}[leftmargin=*]
		\item[(a)] All paths from $w_i$ to $w_\Jc$, excluding the direct path $G_{ji}$, pass through a node $w_k, k \in \Kc \setminus \{i\}$,
		\item[(b)] All paths from $w_j$ to $w_\Jc$ pass through a node $w_k, k \in \Kc$.
\end{enumerate}
\end{thm}
\textbf{Proof:} Collected in the appendix.
\hfill $\Box$

In condition \textit{(a)} the index $i$ is excluded from the set $\Kc$ since every path that starts in $w_i$ contains a node in $\Kc$.
Conditions \emph{(a)} and \emph{(b)} imply that there cannot be any direct paths from $w_i$ and $w_j$ to indirect observations $w_\Lc$, i.e. $G_{\Lc i} = 0$, and $G_{\Lc j} = 0$.

The set $\Kc$ is the set of, either directly retained signals in $\tilde\Sc$, except for node $j$, or indirectly observed nodes in $\Vc$.
The result of the theorem implies that all parallel paths from $w_i$ to $w_\Jc$ and all loops around the 'output', i.e. all paths from $w_j$ to $w_\Jc$, must pass through a node in this set.
\begin{rem}
	The conditions in Theorem \ref{prop:cond_gen} are a generalization of the conditions for  immersion.
	For the choice $\Lc= \emptyset$ and $\Vc = \emptyset$, Algorithm \ref{alg:gen} is equivalent to the immersion algorithm, and the results of Theorem \ref{prop:cond_gen} are equivalent to the conditions of Proposition \ref{prop.setD}, \citep{dankers2016identification}.
	In the generalized situation, parallel paths $w_i \rightarrow w_j$ and loops around the output $w_j \rightarrow w_j$ can also be blocked by indirectly observed nodes, present in $\Vc$, instead of just by directly observed nodes in $\tilde\Sc$.
\end{rem}
\begin{rem}
	The conditions in Theorem \ref{prop:cond_gen} are a generalization of the conditions for the indirect inputs method, as formulated in \cite{linder2017identification}. This latter method results if we consider the particular situation that indirect observations are no in-neighbors of the output node, i.e. $G_{j\Lc}=0$, and that all in-neighbors of indirect observations are in $\Sc \cup \Vc$, i.e. $G_{\Lc\tZc}=0$.
In the generalized situation presented here, indirect observations $w_\Lc$ are allowed to be in-neighbors of $w_j$, and they are allowed to have abstracted nodes $w_\Zc$ as in-neighbors.
\end{rem}
\section{Node selection strategy} \label{sect:identsetup}
Theorem \ref{prop:cond_gen} allows us to check whether a module remains invariant under abstraction if the network topology is known and we have divided the nodes into four groups.
The next question is how to choose the sets of nodes, based on the network topology, such that the module of interest remains invariant.
\subsection{Selecting the sets of nodes} \label{sect:choose}
The strategy to obtaining a set of measured nodes in \cite{dankers2016identification} is as follows.
First the input and output nodes of the module of interest are required to be measured.
Then every parallel path from the input to the output node must be blocked by a measured node.
This means that nodes are added such that each of those paths contains a measured node.
Similarly every loop around the output node must be blocked by a measured node, so nodes are added such that each of those loops contains a measured node.
Different nodes on a path can be chosen to block the path, so the choice of which nodes to measure is not unique.

Now, the method of choosing nodes is adapted with the possibility of using indirectly observed nodes.
A parallel path or a loop can now be blocked by either a measured or an indirectly observed node.
However, when we use an indirect observation to block a path, additional conditions must be satisfied.
Paths from either input or output of the module of interest to the indirect observation must also be blocked by either a measured or an indirectly observed node.
For each indirect observation that is added, this condition on blocking the paths is applied recursively.
This selection method is demonstrated in the following example.
\begin{exmp}[Selecting nodes]\label{ex:nodeselect}
	\begin{figure}[tbh]
		\centering
		\includegraphics[width=0.6\columnwidth]{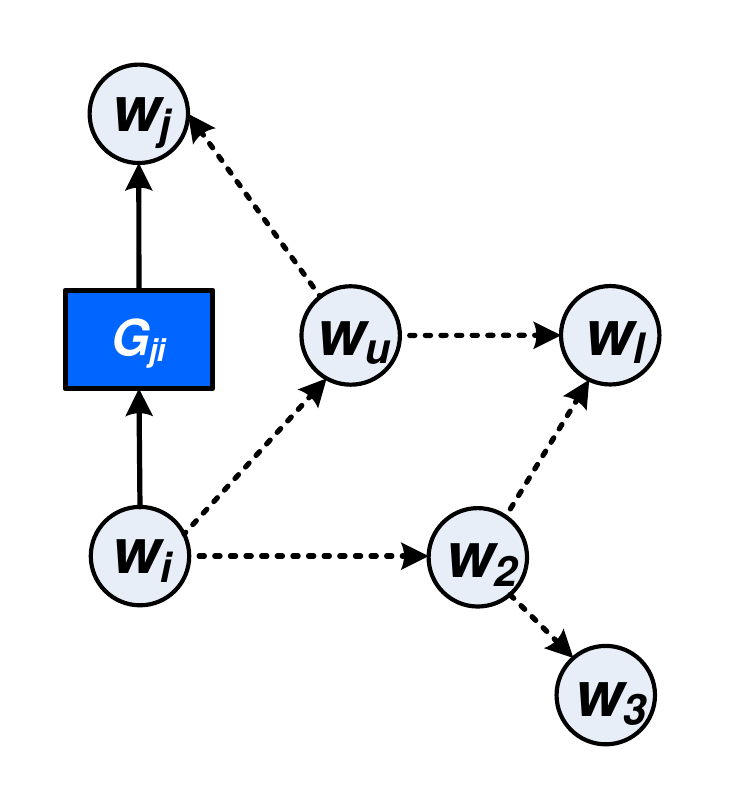}
		\caption{Network where measured nodes and indirectly observed nodes are to be selected.}
		\label{fig:select}
	\end{figure}
	For an illustration of how to select nodes, consider the network in Figure \ref{fig:select}.
	The module of interest is $G_{ji}$, so we select $w_j$ as output, and $w_i$ is included as a predictor input.
	A parallel path through node $w_u$ exists and must be blocked if $G_{ji}$ is to remain invariant.
	We can either include $w_u$ as a predictor input, or we can choose to indirectly observe it using $w_l$.
	When $w_l$ is chosen as indirect input measurement, $l \in \Lc$, and the parallel path from $w_i$ to $w_l$ through $w_2$ should be blocked, so either $w_2$ should then be included as a predictor input, or  $w_3$ can be included as the indirect observation of $w_2.$ \hfill $\square$
\end{exmp}
\subsection{External excitation}
In an estimation setting, both the nodes  $ w_k, k \in \tSc\setminus j$ and $w_l, l \in \Lc$ are used as predictor inputs to parameterized modules.
The question can then be raised what the effect is on the identification setup for nodes being present in one of these sets.
The external variables that are in-neighbors of the node $w_j$ in the abstracted network need to be included as input.
Depending on the chosen nodes in $\tSc$ and $\Lc$ different external variables are  in-neighbors of the node $w_j$, so different external variables need to be chosen in the experimental setup.
We have seen that placing a node in either $\Lc$ or $\tSc$ leads to a different transformation matrix $P^{(abs)}$.
This leads to different locations of zeros in the transformed $\check R$.
The structure of  $\check R$ can be described as follows.
Let $D$ denote the structure of a matrix that is diagonal, and let $\ast$ denote a matrix of arbitrary structure, then
\begin{equation} \label{eq:ustruct}
 	\check u_\Sc   =
 	\begin{bmatrix}
 		  D & \ast & 0 & \ast
 		\\
 		0 & \ast & D \; G_{\Lc \Vc}^{(1)} & \ast
 	\end{bmatrix}   R   \begin{bmatrix}
 		   r_{\tilde \Sc}    \\    r_\Lc    \\    r_\Vc    \\    r_{\tilde \Zc}
 	\end{bmatrix}      ,
\end{equation}
where
\begin{equation} \label{eq:Glv1}
 	G_{\Lc \Vc}^{(1)}  = \left (G_{\Lc \Vc} + G_{\Lc \tilde \Zc}(I-G_{\tilde \Zc \tilde \Zc})^{-1}G_{\tilde \Zc \Vc} \right ).
\end{equation}
If we consider situation that the conditions of Proposition \ref{prop:diagR} are satisfied, e.g. where  $R$ and  $G_{\Lc \Vc}$ are diagonal, where $G_{\Lc\tZc}=0$, $G_{\Vc\Vc}=0$, and $G_{\Vc\tZc}=0$, and considering that $D$ is diagonal,
then the external excitations
\[  r_j,	\quad	r_\Lc, \quad \text{ and }	r_{\tZc}  \]
may be in-neighbor of $w_j$.
In terms of choosing a network model set for the abstracted network, the structure of \eqref{eq:ustruct} specifies how to choose the zero-structure of the parameterized $\check R(q,\theta)$ that is to be used for estimation of the abstracted network.
\subsection{The noise model}
Due to the abstraction, the noise process is modified  in a way that is the same as the modification of  the external excitations.
The following expression is obtained for the disturbances
\begin{equation} \label{eq:vstruct}
 	\check v_\Sc    =
 	\begin{bmatrix}
 		   D & \ast & 0 & \ast
 		\\
 		0 & \ast & D \; G_{\Lc \Vc}^{(1)}  & \ast
 	\end{bmatrix}    H    \begin{bmatrix}
 		   e_{\tilde \Sc}    \\    e_\Lc    \\    e_\Vc    \\    e_{\tilde \Zc}
 	\end{bmatrix}      .
\end{equation}
with $G_{\Lc \Vc}^{(1)} $ specified in \eqref{eq:Glv1}.
The obtained noise filter above is not square, which is problematic in terms of identification.
This noise model relates to a square noise filter $\check H$ and white noise $\check e$ that can be used in an identification setting.
It is likely that the obtained  $\check H$  is then no longer diagonal.
The zero-structure of the obtained noise filter can be used as the zero-structure when parameterizing the network model set.
Under particular conditions special noise structures can be obtained that can be exploited.
If no particular structure is obtained for the noise model, then all process noises are correlated.
%
%
%
\subsection{Identification methods}
For a particular network, and a choice of target module $G_{ji}$,  the choice of the node sets $\tSc,\Lc,\Vc,\tZc$ will determine whether the target module will remain invariant in the abstracted network.
This result can be applied in the problem of identifying the target module $G_{ji}$ on the basis of measured node signals.
In the abstracted network we have the node signals $w_{\Sc}$, which we assume to be available from observations.
We can now construct an identification setup in line with the methods developed in \cite{VandenHof&Dankers&etal:13}.
Determine the set of input predictors as those node signals in $w_{\Sc}$ that are in-neighbors of the output $w_j$ in the abstracted network.
If $G_{ji}$ has remained invariant in the abstracted network, i.e. when the conditions of Theorem \ref{prop:cond_gen} are satisfied, he identification problem of estimating the transfer functions from inputs in $w_{\Sc}$ to output $w_j$ will now estimate the module from input $w_i$ to output $w_j$ that is equal to the module $G_{ji}$ in the original network.
Consistent identification of this module is then possible under the typical regularity conditions of the prediction error methods, as formulated in \cite{VandenHof&Dankers&etal:13}.
This implies that:
\begin{itemize}
\item For the two-stage identification method, consistency of the estimate $\hat G_{ji}$ is achievable if there is a sufficient excitation by external excitation signals in the network;
\item For the direct identification method, consistency of the estimate $\hat G_{ji}$ is achievable, if besides sufficient excitation by external excitation and disturbance signals, correlated noises between inputs and outputs are taken care of. This can be done by either choosing the node sets $\tSc,\Lc,\Vc,\tZc$ such that these correlated noises (or confounding variables) do not occur, \citep{dankers2017conditions}, or by modeling this noise correlation correctly in the model, leading to the so-called joint-direct method \citep{weerts2018rr,VandenHof&etal_CDC:19}.
\end{itemize}
When applied to the abstracted network, the prediction error associated with the joint-direct method is
\begin{equation}
 	\varepsilon(\theta) = \check H^{-1}(\theta)
 	\left (
 		(I-\check G(\theta))w_\Sc - \check R(\theta) r
 	\right ),
\end{equation}
where the $\check G(\theta), \check R(\theta), \check H^{-1}(\theta) $ are structured according to the topology obtained by network abstraction, the $w_\Sc$ are all retained nodes, and $r$ are all available external excitations.
Then $\varepsilon^T \varepsilon$ is minimized over the parameters to obtain the estimated model.
A further analysis of the particular identification results is beyond the scope of this paper.
\section{Conclusions}
The question to be answered is which set of measured nodes can lead to consistent estimates of a target module.
As a way to answer this question the concept of abstraction has been introduced as a way to remove unmeasured nodes from a network representation, as a generalization of methods present in literature.
A systematic method has been introduced to select nodes such that the module of interest remains invariant in the abstracted network.
Under some assumptions on external excitations and the network topology the abstracted network can be parameterized with a network identifiable model set.
If the module of interest remains invariant, and the model set is identifiable, then conditions for consistent estimation can be obtained for various identification methods.

A requirement that has been imposed is that the module of interest remains invariant in the abstracted network, but this is not necessary for consistency.
It may be that the module of interest can be identified in an indirect way by combining the knowledge of two or more modules present in the abstracted network.
It is also possible that there are multiple sets of measured nodes that each lead to consistent estimates of a module of interest.
Selecting the set of nodes that leads to the smallest variance is another question for future consideration.
\section*{Acknowledgements}
The authors are grateful for the discussions with Arne Dankers on the topic of the paper.
\section{Appendix}
\subsection{Proof of Proposition \ref{prop:transP}}
Sufficiency:
By \eqref{eq:transfo} the diagonal of $G^{(2)}$ is
\begin{equation}
 	\text{diag}\left(G^{(2)}\right) = \text{diag}\left( I - P(I-G^{(1)})\right),
\end{equation}
which is 0 by condition \emph{(2)}, showing that $G^{(2)}$ is hollow. Moreover if Condition \emph{(1)} is satisfied, then
with (\ref{eq:transfo}), (\ref{eq:transfoR}):
\begin{equation}
 	(I-G^{(2)})^{-1}R^{(2)} = (I-G^{(1)})^{-1}R^{(1)}
\end{equation}
which is proper and stable by Definition \eqref{def:netw_model}.
A monic, proper, stable, and inversely stable $H^{(2)}$ and full rank $\Lambda^{(2)}$ are obtained through the spectral factorization in \eqref{eq:transfoH}.\\
Necessity:
In order for $(I-G^{(2)})^{-1}R^{(2)}$ to be proper and be stable, it is required that $P^{-1}$ exists. Therefore $P$ has to have full rank. In order for $G^{(2)}$ to be hollow, it is required that $diag(I-P(I-G^{(1)})) = 0$.
 \hfill $\Box$
\subsection{Proof of Proposition \ref{prop:freeG}}
Substituting $P=(I-G^{(2)})(I-G^{(1)})^{-1}$ into the definition of the transformation \eqref{eq:transfo} gives
\begin{equation}
 	G^{(2)} = I - (I-G^{(2)})(I-G^{(1)})^{-1} (I-G^{(1)}),
\end{equation}
which shows that $G^{(2)}$ is obtained by applying this transformation.
Moreover the diagonal of $(I-P(I-G^{(1)}))$ is 0, so $P=(I-G^{(2)})(I-G^{(1)})^{-1}$ is an appropriate transformation.
\hfill $\Box$
\subsection{Proof of Proposition \ref{prop:eq_alg_trans}}
In order to prove the proposition we evaluate the expressions for each step of Algorithm \ref{alg:gen}.

\textbf{Step a:}
The fourth equation of \eqref{eq:netw_4group} is solved for $w_{\tilde \Zc}$
			\begin{equation} \label{eq:tUc} \begin{split}
			 	 w_{\tilde \Zc} = &(I- G_{\tilde \Zc \tilde \Zc})^{-1} \\ &
			 	 \left (
			 	 G_{\tilde \Zc \tilde \Sc} w_{\tilde \Sc}+G_{\tilde \Zc \Lc}w_\Lc +G_{\tilde \Zc \Vc}w_\Vc
			 	+ u_{\tilde \Zc}
			 	+v_{\tilde \Zc}   \right ) .
			\end{split} \end{equation}
Substituting the above equation into the remainder of the network results for $w_{\tSc}$ in
\begin{equation} \label{eq:s}
	w_{\tSc} = G_{\tSc \tSc}^{(1)} w_{\tSc}  + G_{\tSc \Lc}^{(1)} w_{\Lc} + G_{\tSc \Vc}^{(1)} w_{\Vc} + u_{\tSc}^{(1)} + v_{\tSc}^{(1)},
\end{equation}
with
\begin{align*}
	G_{\tSc \tSc}^{(1)} &= \left ( G_{\tSc \tSc}  +  G_{\tSc \tZc} (I- G_{\tZc \tZc})^{-1}  G_{\tZc  \tSc} \right )
	\\
	G_{\tSc \Lc}^{(1)} &= \left ( G_{\tSc \Lc}  +  G_{\tSc  \tZc} (I- G_{\tZc  \tZc})^{-1}  G_{\tZc \Lc} \right )
	\\
	G_{\tSc \Vc}^{(1)} &= \left ( G_{\tSc \Vc}  +  G_{\tSc  \tZc} (I- G_{\tZc  \tZc})^{-1}  G_{\tZc \Vc} \right )
	\\
	u_{\tSc}^{(1)} &= u_{\tSc} +  G_{\tSc \tZc} (I- G_{\tZc \tZc})^{-1} u_{\tZc}
	\\
	v_{\tSc}^{(1)} &= v_{\tSc} +  G_{\tSc \tZc} (I- G_{\tZc \tZc})^{-1} v_{\tZc}.
\end{align*}
For $w_\Lc$ we obtain
\begin{equation} \label{eq:l}
	w_{\Lc} = G_{\Lc \tSc}^{(1)} w_{\tSc}  + G_{\Lc \Lc}^{(1)} w_{\Lc} + G_{\Lc \Vc}^{(1)} w_{\Vc} + u_{\Lc}^{(1)} + v_{\Lc}^{(1)},
\end{equation}
with
\begin{align*}
	G_{\Lc \tSc}^{(1)}& = \left ( G_{\Lc \tSc}  +  G_{\Lc \tZc} (I- G_{\tZc \tZc})^{-1}  G_{\tZc  \tSc} \right )
	\\
	G_{\Lc \Lc}^{(1)}& = \left ( G_{\Lc \Lc}  +  G_{\Lc  \tZc} (I- G_{\tZc  \tZc})^{-1}  G_{\tZc \Lc} \right )
	\\
	G_{\Lc \Vc}^{(1)} &= \left ( G_{\Lc \Vc}  +  G_{\Lc  \tZc} (I- G_{\tZc  \tZc})^{-1}  G_{\tZc \Vc} \right )
	\\
	u_{\Lc}^{(1)}& = u_{\Lc} +  G_{\Lc \tZc} (I- G_{\tZc \tZc})^{-1} u_{\tZc}
	\\
	v_{\Lc}^{(1)} &= v_{\Lc} +  G_{\Lc \tZc} (I- G_{\tZc \tZc})^{-1} v_{\tZc}.
\end{align*}
For $w_\Vc$ we obtain
\begin{equation} \label{eq:v}
	w_{\Vc} = G_{\Vc \tSc}^{(1)} w_{\tSc}  + G_{\Vc \Lc}^{(1)} w_{\Lc} + G_{\Vc \Vc}^{(1)} w_{\Vc} + u_{\Vc}^{(1)} + v_{\Vc}^{(1)},
\end{equation}
with
\begin{align*}
	G_{\Vc \tSc}^{(1)} &= \left ( G_{\Vc \tSc}  +  G_{\Vc \tZc} (I- G_{\tZc \tZc})^{-1}  G_{\tZc  \tSc} \right )
	\\
	G_{\Vc \Lc}^{(1)} &= \left ( G_{\Vc \Lc}  +  G_{\Vc  \tZc} (I- G_{\tZc  \tZc})^{-1}  G_{\tZc \Lc} \right )
	\\
	G_{\Vc \Vc}^{(1)} &= \left ( G_{\Vc \Vc}  +  G_{\Vc  \tZc} (I- G_{\tZc  \tZc})^{-1}  G_{\tZc \Vc} \right )
	\\
	u_{\Vc}^{(1)} & = u_{\Vc} +  G_{\Vc \tZc} (I- G_{\tZc \tZc})^{-1} u_{\tZc}
	\\
	v_{\Vc}^{(1)} &= v_{\Vc} +  G_{\Vc \tZc} (I- G_{\tZc \tZc})^{-1} v_{\tZc}.
\end{align*}
It is straightforward to verify that the transformation
$ G^{(1)} = I-P^{(1)}(I-G)$, $u^{(1)} =P^{(1)}u$, and $v^{(1)} =P^{(1)}v$
results in the same expressions as the algorithm.

\textbf{Steps b and c:}
The two equations that are solved for $w_\Vc$ result in
\begin{equation} \label{eq:v1}
 	w_\Vc = (G_{\Lc \Vc}^{(1)})^\dagger \left ( - G_{\Lc \tSc}^{(1)} w_{\tSc} +(I- G_{\Lc \Lc}^{(1)} ) w_{\Lc} - u_{\Lc}^{(1)} - v_{\Lc}^{(1)} \right ),
\end{equation}
and
\begin{equation}\label{eq:v2}
 	w_\Vc = (I-G_{\Vc \Vc}^{(1)} )^{-1} \left ( G_{\Vc \tSc}^{(1)} w_{\tSc}  + G_{\Vc \Lc}^{(1)} w_{\Lc} + u_{\Vc}^{(1)} + v_{\Vc}^{(1)} \right ).
\end{equation}
Substituting \eqref{eq:v1} into \eqref{eq:s} results in
\begin{equation} \label{eq:s3}
	w_{\tSc} = G_{\tSc \tSc}^{(3)} w_{\tSc}  + G_{\tSc \Lc}^{(3)} w_{\Lc} + u_{\tSc}^{(3)} + v_{\tSc}^{(3)},
\end{equation}
with
\begin{align*}
 	G_{\tSc \tSc}^{(3)} &= G_{\tSc \tSc}^{(1)} - G_{\tSc \Vc}^{(1)} (G_{\Lc \Vc}^{(1)})^\dagger G_{\Lc \tSc}^{(1)}
 	\\
 	G_{\tSc \Lc}^{(3)} &= G_{\tSc \Lc}^{(1)} + G_{\tSc \Vc}^{(1)} (G_{\Lc \Vc}^{(1)})^\dagger (I- G_{\Lc \Lc}^{(1)})
 	\\
	u_{\tSc}^{(3)} &= u_{\tSc}^{(1)} +  G_{\tSc \Vc}^{(1)} (G_{\Lc \Vc}^{(1)})^\dagger u_{\Lc}^{(1)}
	\\
	v_{\tSc}^{(3)} &= v_{\tSc}^{(1)} +  G_{\tSc \Vc}^{(1)} (G_{\Lc \Vc}^{(1)})^\dagger v_{\Lc}^{(1)} .
\end{align*}
Substituting \eqref{eq:v1} into \eqref{eq:l} results in
\begin{equation} \label{eq:l3}
	w_{\Lc} = G_{\Lc \tSc}^{(3)} w_{\tSc}  + G_{\Lc \Lc}^{(3)} w_{\Lc} + u_{\Lc}^{(3)} + v_{\Lc}^{(3)},
\end{equation}
with
\begin{align*}
	G_{\Lc \tSc}^{(3)} &= G_{\Lc \tSc}^{(1)} + G_{\Lc \Vc}^{(1)} (I-G_{\Vc \Vc}^{(1)} )^{-1}  G_{\Vc \tSc}^{(1)}
	\\
	G_{\Lc \Lc}^{(3)}  &= G_{\Lc \Lc}^{(1)}  + G_{\Lc \Vc}^{(1)} (I-G_{\Vc \Vc}^{(1)} )^{-1} G_{\Vc \Lc}^{(1)}
	\\
	u_{\Lc}^{(3)} &= u_{\Lc}^{(1)} +  G_{\Lc \Vc}^{(1)} (I-G_{\Vc \Vc}^{(1)} )^{-1} u_{\Vc}^{(1)}
	\\
	v_{\Lc}^{(3)} &= v_{\Lc}^{(1)} +  G_{\Lc \Vc}^{(1)} (I-G_{\Vc \Vc}^{(1)} )^{-1} v_{\Vc}^{(1)} .
\end{align*}
The combination of the transformations $P^{(3)} P^{(2)}$ is denoted by
\[ P^{(3,2)} :=
\begin{bmatrix}
	I & G^{(2)}_{\tilde \Sc \Vc} (G_{\Lc\Vc}^{(1)})^\dagger &0 & 0 \\
	0 & I & G_{\Lc\Vc}^{(1)}(I-G_{\Vc\Vc}^{(1)})^{-1} & 0 \\
	0 & (G_{\Lc\Vc}^{(1)})^\dagger & 0 & 0 \\
	0 & G^{(2)}_{\tilde \Zc \Vc} (G_{\Lc\Vc}^{(1)})^\dagger  & 0 & I
\end{bmatrix},
\]
and it is straightforward to verify that this transformation
$ G^{(3)} = I-P^{(3,2)}(I-G^{(1)})$,  $u^{(3)} =P^{(3,2)}u^{(1)}$, and $v^{(3)} =P^{(3,2)}v^{(1)}$
results in the same expressions as the algorithm.

\textbf{Step d:}
Removal of a self-loop in the equation for node $w_j$ is performed by subtracting the right-hand term for $w_j$ from both sides, and then dividing both sides by 1 minus the term.
This is precisely the operation performed by the transformation $P^{(4)}$.
\hfill $\Box$
\subsection{Proof of Proposition \ref{prop:diagR}}
Transfer function matrix $\check R$ is
\begin{equation}
 	\check R =  \begin{bmatrix}	I&0	 \end{bmatrix} P^{(4)}P^{(3)}P^{(2)}P^{(1)} R,
\end{equation}
where the transformations are defined in Section \ref{sect:trans}.
The structural properties of the matrices will be evaluated.
Let $\ast$ indicate an unstructured matrix, and let $D$ indicate a diagonal matrix structure.
Then the structure of the first part of the transformation matrix is
\begin{equation}
 	\begin{bmatrix}	I&0	 \end{bmatrix} P^{(4)}P^{(3)} =
 	\begin{bmatrix}	I&0	 \end{bmatrix}
 	\begin{bmatrix}
 		D & 0 \\ 0 & D
 	\end{bmatrix}
 	\begin{bmatrix}
 		D&\ast \\ 0 & \ast
 	\end{bmatrix}
 	=
 	 \begin{bmatrix}
 		D&\ast
 	\end{bmatrix},
\end{equation}
where the matrices are partitioned corresponding to the blocks  $\Sc = \tilde \Sc \cup \Lc$ and  $\Zc = \tilde \Zc \cup \Vc$.\\
The transformation $P^{(2)}P^{(1)}$ has the following structure
\begin{equation}
 	P^{(2)}P^{(1)}
 	=
 	\begin{bmatrix}
 		D&0&0&0 \\ 0 & D&X&0 \\ 0& \ast&0&0 \\ 0&0&0&D
 	\end{bmatrix}
 	\begin{bmatrix}
 		D&0&0&\ast \\ 0 & D&0&\ast \\ 0& 0&D&\ast \\ 0&0&0&\ast
 	\end{bmatrix}
 	=
 	\begin{bmatrix}
 		D&0&0&\ast \\ 0 & D&X&\ast \\ 0& \ast&0&\ast \\ 0&0&0&\ast
 	\end{bmatrix},
\end{equation}
where the matrices are partitioned corresponding to the blocks $\tSc,\Lc,\Vc,\tZc$, and where
$X = G_{\Lc\Vc}^{(1)}(I-G_{\Vc\Vc}^{(1)})^{-1}$.
From the relations in  \eqref{eq:l} and \eqref{eq:v}
we obtain that
\[  G_{\Lc\Vc}^{(1)} = \left ( G_{\Lc \Vc}  +  G_{\Lc  \tZc} (I- G_{\tZc  \tZc})^{-1}  G_{\tZc \Vc} \right )\]
and
\[ G_{\Vc\Vc}^{(1)} = \left ( G_{\Vc \Vc}  +  G_{\Vc  \tZc} (I- G_{\tZc  \tZc})^{-1}  G_{\tZc \Vc} \right ).\]
Then from condition 3 we obtain that $X$ is diagonal.

Under conditions 1 and 2, matrix $R$ has the structure
\begin{equation}
 	R = \begin{bmatrix}
 		D&\ast&0&\ast \\ 0&\ast&0&\ast \\ 0&\ast&D&\ast \\ 0&\ast&0&\ast
 	\end{bmatrix},
\end{equation}
such that the following is obtained
\begin{equation}
 	P^{(2)}P^{(1)}R = \begin{bmatrix}
 		D&\ast&0&\ast \\ 0&\ast&D&\ast \\ 0&\ast&0&\ast \\ 0&\ast&0&\ast
 	\end{bmatrix}.
\end{equation}
Then the final structure is
\begin{equation}
 	\check R =
 	\begin{bmatrix}
 		D & 0 & \ast &\ast
 		\\
 		0 & D & \ast &\ast
 	\end{bmatrix}
 	\begin{bmatrix}
 		D&\ast&0&\ast \\ 0&\ast&D&\ast \\ 0&\ast&0&\ast \\ 0&\ast&0&\ast
 	\end{bmatrix}
 	=
 	\begin{bmatrix}
 		D&\ast&0&\ast \\ 0&\ast&D&\ast  		
 	\end{bmatrix}.
\end{equation}
It is then obvious that a leading diagonal can be obtained by column operations.
\hfill $\Box$
\subsection{Proof of Theorem \ref{prop:cond_gen}}
In order to prove the theorem, the conditions must be interpreted in terms of $G$.
Conditions \emph{(a)} and \emph{(b)} imply that there are no direct paths from $w_i$ and $w_j$ to indirect observations $w_\Lc$, i.e.
\begin{enumerate}
	\item[\emph{(i)}] $G_{\Lc i} = 0$,
	\item[\emph{(ii)}] $G_{\Lc j} = 0$.
\end{enumerate}
The conditions also imply that there are no paths from $w_i$ and $w_j$ to indirect observations $w_\Lc$ and $j$ that only go through unmeasured nodes $w_{\tZc}$, i.e.
\begin{enumerate}
	\item[\emph{(iii)}] $G_{\Lc\tZc}(I-G_{\tZc\tZc})^{-1}G_{\tZc i} = 0$,
	\item[\emph{(iv)}] $G_{\Lc\tZc}(I-G_{\tZc\tZc})^{-1}G_{\tZc j}  = 0$,
	\item[\emph{(v)}] $G_{j \tZc}(I-G_{\tZc\tZc})^{-1}G_{\tZc i}  = 0$,
	\item[\emph{(vi)}] $G_{j \tZc}(I-G_{\tZc\tZc})^{-1}G_{\tZc j}  = 0$.
\end{enumerate}
The module of interest is a part of $G^{(4)}$ in \eqref{eq:G4} which can be obtained as
\begin{equation}
 	G^{(4)} = I-P^{(4)}(I-G^{(3)}).
\end{equation}
Explicit expressions can be found in the proof of Proposition \ref{prop:eq_alg_trans} such that using \emph{(i)-(vi)} we can see that $G^{(4)}_{ji} = G_{ji}$.
First it is shown that $G^{(3)}_{ji} = G_{ji}$.
From \eqref{eq:s3} we obtain that
\begin{equation}
 	G^{(3)}_{ji} = G_{ji}^{(1)} - G_{j \Vc}^{(1)} (G_{\Lc \Vc}^{(1)})^\dagger G_{\Lc i}^{(1)}.
\end{equation}
Then it can be observed that
\begin{equation}
 	G_{\Lc i}^{(1)} = G_{\Lc i}  +  G_{\Lc \tZc} (I- G_{\tZc \tZc})^{-1}  G_{\tZc  i} = 0
\end{equation}
if we use the expression from \eqref{eq:l} and the obtained conditions \emph{(i)} and \emph{(iii)}.
Now $G^{(3)}_{ji} = G_{ji}^{(1)}$ is evaluated using the expression in \eqref{eq:s}
\begin{equation}
 	G_{ji}^{(3)} = G_{ji}  +  G_{j \tZc} (I- G_{\tZc \tZc})^{-1}  G_{\tZc  i} .
\end{equation}
Then by condition \emph{(v)} we have $G_{ji}^{(3)} = G_{ji}$.

Since $P^{(4)}$ is diagonal, all that is left to show is that its $jj$-th entry is 1.
Using the expression \eqref{eq:P4} we then need to show that $G_{jj}^{(3)}=0$.
From \eqref{eq:s3} we obtain that
\begin{equation}
 	G^{(3)}_{jj} = G_{jj}^{(1)} - G_{j \Vc}^{(1)} (G_{\Lc \Vc}^{(1)})^\dagger G_{\Lc j}^{(1)}.
\end{equation}
Then it can be observed that
\begin{equation}
 	G_{\Lc j}^{(1)} = G_{\Lc j}  +  G_{\Lc \tZc} (I- G_{\tZc \tZc})^{-1}  G_{\tZc  j} = 0
\end{equation}
if we use the expression from \eqref{eq:l} and the obtained conditions \emph{(ii)} and \emph{(iv)}.
Now $G_{jj}^{(1)}$ is evaluated using the expression in \eqref{eq:s}
\begin{equation}
 	G_{jj}^{(1)} = G_{jj}  +  G_{j \tZc} (I- G_{\tZc \tZc})^{-1}  G_{\tZc  j} .
\end{equation}
Since there are no self loops in $G$, the $G_{jj} =0$.
Then by condition \emph{(vi)} we have $G_{jj}^{(3)} = 0$,
such that $P^{(4)}_{jj}=1$ and $G_{ji}^{(4)} = G_{ji}$.
\hfill $\Box$
\bibliographystyle{plainnat}        
\bibliography{Library}           
%
%
\end{document}